# Significant reduced traffic in Beijing failed to relieve haze pollution during the COVID-19 lockdown: implications for haze mitigation

Zhaofeng Lv[a,1], Xiaotong Wang [a,1], Fanyuan Deng [a], Qi Ying[b], Alexander T. Archibald[c], Roderic L. Jones[c], Yan Ding[d], Ying Cheng[e], Mingliang Fu[d], Ying Liu[e], Hanyang Man[a], Zhigang Xue[d], Kebin He [a], Jiming Hao[a] & Huan Liu [a,2]

[a]State Key Joint Laboratory of ESPC, State Environmental Protection Key Laboratory of Sources and Control of Air Pollution Complex, International Joint Laboratory on Low Carbon Clean Energy Innovation, School of the Environment, Tsinghua University, Beijing 100084, China; [b]Zachry Department of Civil and Environmental Engineering, Texas A&M University, College Station, TX 77843, USA; [c]Centre for Atmospheric Science, Department of Chemistry, University of Cambridge, CB2 1EW, UK; [d]Chinese Research Academy of Environmental Sciences, Beijing 100012, China; [e]Beijing Transport Institute, Beijing 100073, China

[1]Z. Lv, and X. Wang contributed equally to this work.
[2]Correspondence to Huan Liu.

Email: liu_env@tsinghua.edu.cn.

The COVID-19 outbreak greatly limited human activities and reduced primary emissions particularly from urban on-road vehicles, but coincided with Beijing experiencing "pandemic haze", raising the public concerns of the validity and effectiveness of the imposed traffic policies to improve the air pollution. Here, we explored the relationship between local vehicle emissions and the winter haze in Beijing before and during the COVID-19 lockdown period based on an integrated analysis framework, which combines a real-time on-road emission inventory, in-situ air quality observations and a localized chemical transport modeling system. We found that traffic emissions decreased substantially affected by the pandemic, with a higher reduction for $NO_x$ (75.9%, 125.3 Mg/day) compared to VOCs (53.1%, 52.9 Mg/day). Unexpectedly, our results show that the imbalanced emission abatement of $NO_x$ and VOCs from vehicles led to a significant rise of the atmospheric oxidizing capacity in urban areas, but only resulting in modest increases in secondary aerosols due to the inadequate precursors. However, the enhanced oxidizing capacity in the surrounding regions greatly increased the secondary particles with relatively abundant precursors, which is mainly responsible for Beijing haze during the lockdown period. Our results indicate that the winter haze in Beijing was insensitive to the local vehicular emissions reduction due to the complicated nonlinear response of the fine particle and air pollutant emissions. We suggest mitigation policies should focus on accelerating VOC and $NH_3$ emissions reduction and synchronously controlling regional sources to release the benefits on local traffic emission control.

The unexpected COVID-19 epidemic in 2020, which coincided with the Spring Festival, the most important holiday in China, put the Chinese economy into a rapid stall. The Spring Festival migration reduced the population in Beijing to a low level, with an estimated 39% reduction from the 22 million residents in normal times (1), while the Coronavirus pandemic lockdown further reduced activities. The stay-at-home orders were initially started in Wuhan from January 23rd, 2020, one day before the Eve of the Spring Festival, and soon applied to the whole country (2). In addition, for those who enter Beijing, a two-week compulsory quarantine was implemented. The Spring Festival holiday and the Coronavirus restrictions led to widespread shutdowns and a near-halt in normal life and economic activity in Beijing and its surrounding cities. Generally, pandemic lockdowns led to clearer skies in China and other places in the world (3-5). However, severe air pollution episodes occurred in Beijing during "the most silent spring", leading to query the response of air pollution to anthropogenic activities.

The unexpected heavy pollution put doubt on current understanding of the source-receptor relationship in Beijing. Previous work demonstrated that vehicles contributed 45% to ambient $PM_{2.5}$ caused by all local sources (6). Following scientific instructions, control measurements were undertaken over the past 5-years to reduce the sources of aerosol pollution (7). As a result, the $PM_{2.5}$ annual concentration in Beijing decreased from 89.5 μg/m$_3$ in 2013 to 42.0 μg/m$_3$ in 2019, and heavy pollution days were also reduced from 58 days in 2013 to 4 days in 2019 (8), providing confidence in source-receptor mechanisms supporting pollution control strategies. However, during the COVID-19 lockdown the air quality index (AQI) frequently hit extremely unhealthy levels in Beijing, including January 25$_{th}$-29$_{th}$ and February 9$_{th}$-13$_{th}$, with a peak daily $PM_{2.5}$ concentration reaching 218.3 μg/m$_3$ on February 12$_{th}$, more than 8 times higher than the World Health Organization (WHO)'s recommended level of 25 μg/m$_3$ for 24-hr average concentrations. Here we show that the severe haze pollution, which occurred in spite of the significantly reduced human activities, highlights weaknesses in our current source-receptor understanding.

The "pandemic haze" in Beijing raised great attention from the public and government on the role of vehicle emissions in the burden of air pollution. Historically, vehicle emission controls have been used as an effective way to relieve air pollution in megacities (9, 10). Beijing undertook a lot of effort to reduce its traffic emissions through strict controls on new vehicle registration, limited usage based on plate number, upgraded vehicle emission standards and shifting to greener transportation (11, 12). In Beijing, car license plate is regarded as a limited public resource. To get a conventional gasoline car, residents need to be in a bimonthly lottery pool, competing with more than 3 million fellow residents with an odds of around 1/2900. Beijing residents believed these restrictions could help improve the air quality, as official reports showed that the vehicles contributed 45% to ambient $PM_{2.5}$ from local sources in Beijing (6). However, recent air pollution episodes in Beijing have made the relationship tangled. What is the role of vehicular emission reduction in $PM_{2.5}$ pollution during the COVID-19 outbreak? Is traffic emission control still a necessary and effective way to relieve the winter haze in a megacity like Beijing?

Here, we presented a source-receptor analysis on the COVID-19 pandemic haze events in Beijing based on emission inventory, air quality observations and numerical models. Our study integrated multiple real-time traffic data around the COVID-19 outbreak and developed a novel realistic traffic emission inventory for Beijing. It was applied in a series of counterfactual modeling experiments by a localized chemical transport modeling system and a tracing-based source apportionment to understand the mechanisms and the role of local vehicle emission reductions for the pandemic haze, and to propose the future development of vehicle emission control strategy. The details of our analytic approach are documented in *Materials and Method* and the *SI Appendix*.

**Traffic activity variations and emission changes during lockdown**

The overall research period was divided into three sub-periods, all occurring in the early spring: the Pre-lockdown period (referred to Pre, before January 20$_{th}$), the Transition period (January 20$_{th}$-23$_{th}$) and the Lockdown period (January 24$_{th}$-February 14$_{th}$, including an overlap from January 24$_{th}$ to 30$_{th}$ of Spring Festival holiday). Multiple high-resolution traffic data were integrated during these periods, including hourly traffic speeds within the 6$_{th}$ ring road for thousands of road links (Fig. 1-A&SI Appendix, Fig. S1), trajectory data from more than one hundred thousand floating cars (vehicles equipped with GPS) (Fig. 1-B) and estimated traffic flow by the newly developed Street-Level On-road Vehicle Emission (SLOVE) model (SI Appendix, section S1 and Fig. S2). Four days before the Spring Festival (Transition period), people began returning to their home towns, and the traffic condition became better with road speeds increased by 5.0%-14.1%. During the COVID-19 Lockdown period, the estimated traffic flow in freeways and urban roads dropped by 37.3%-59.8% compared to the Pre period, with the traffic speed further increased by 14.0%-30.8%, especially for roads within the 5th ring road at morning rush hours. Average daily Vehicle Kilometers of Travel (VTK) of Light Duty Vehicles (LDV), Heavy Duty Vehicles (HDV), Light Duty Trucks (LDT) and Heavy Duty Trucks (HDT) decreased by 28.3%, 60.9%, 36.7% and 55.4%, respectively, during the Lockdown period. After the 7-day Spring Festival holiday, the activity of LDTs gradually increased in order to meet the urban demand, but was still 29.4% lower than the level during the Pre period. Compared with the data of the same period around the Spring Festival in 2019, the traffic speed affected by the pandemic still remained at a high level after the 7-day holiday during the Lockdown period, with 6.3%-13.8% higher than that in 2019 if snowy days were excluded (SI Appendix, Fig. S2). These increases in traffic speed and decreases in traffic flow in Beijing, for such a long time, were significantly more marked than any previous holidays in the past few years (13).

Based on the SLOVE model and TrackATruck model, the real-time on-road emissions of Beijing were calculated for the Pre period, the Transition period, and the Lockdown period. As a consequence of the COVID-19 pandemic, vehicle emissions decreased by 50.9%-75.9%, with particularly high reduction (75.9%) for $NO_x$ emissions (Fig. 1-C, other pollutants were shown in SI Appendix, Fig. S3). The relative reduction of emissions from transportation sector was larger than the averaged decrease of all sectors, which were decreased by 20.6%-69.3% during the Lockdown period. The spatial distribution showed that in the Lockdown period, vehicle emissions decreased substantially on almost all roads, especially for ring roads during the traffic rush hours and the main freight channels at night (SI Appendix. Fig. S4). The diurnal variations of vehicle emissions also showed a significant change during the Lockdown period (SI Appendix. Fig. S5). Two emission peaks were observed on both weekdays and weekends during the Pre period, with the highest hourly $NO_x$ emissions reaching up to 9.2 Mg/hour at 17:00 on weekdays. However, during the Lockdown period, the hourly on-road emissions showed much smaller variations and the difference between weekdays and weekends became smaller. Meanwhile, $NO_x$ emissions at the evening traffic peak declined to 1.8 Mg/hour. In sum, our results indicate that the COVID-19 outbreak led to a significant reduction in traffic activities and emissions compared to those in the Pre period, and also changed the spatial distribution and diurnal variations of vehicle emissions.

The Pre period vehicle emission estimates in this study were at the same magnitude with recent Chinese government led research (SI Appendix, Fig. S6). Before COVID-19 lockdown, on-road emissions were estimated to be 496.8 Mg/day (CO), 99.6 Mg/day (HC), 165.1 Mg/day ($NO_x$), and 5.1 Mg/day ($PM_{2.5}$), accounting for 45.8%, 29.3%, 65.6%, and 21.3% of the total anthropogenic emissions, respectively. HDTs were responsible for only 12.9% and 9.9% of $NO_x$ and $PM_{2.5}$ emissions from all vehicles, much lower than previous estimations (14, 15), because of the implementation of a HDT low emission zone in Beijing. Compared to estimations for 2013 based on a similar bottom-up method (14, 15), $NO_x$ and $PM_{2.5}$ emissions from on-road traffic reduced by 43.5%-49.4% and 52.3%-55.3%, indicating the effectiveness of the continuous vehicular emission control measures. All these features influenced the role of local traffic emissions in air pollution.

**Characteristics of $PM_{2.5}$ pollution indicates its link with local vehicles**

Fig. 2-A shows the observed temporal variations of daily AQI and $PM_{2.5}$ concentrations before and during the COVID-19 outbreak, respectively. The haze pollution became more severe during the Lockdown period compared to that in either the Pre period or the Transition period, with the mean daily $PM_{2.5}$ level unexpectedly increasing from 48.0 μg/m3 to 99.0 μg/m3. Moreover, in the Lockdown period, half of the days were polluted with daily $PM_{2.5}$ concentrations exceeding 75 μg/m3, the level II standard of the Chinese National Ambient Air Quality Standards (NAAQS). The $PM_{2.5}$ level remained at more than 150 μg/m3 for two episodes from January 25th to January 28th and February 11th to 13rd (the first of these was excluded from this analysis since it was probably caused by fireworks (16)). The variation of $PM_{2.5}$ concentrations on polluted days was seen as an asymmetric "saw tooth" pattern, rising slowly before two days and then falling abruptly (17). Both in the Pre period and Lockdown period, these "saw tooth" periods were selected as the heavy pollution periods (HPPs), and other consecutive clean days (daily $PM_{2.5}$ level less than 75 μg/m3) were defined as non-heavy pollution periods (NHPPs). In this study, we separate episodes of HPPs and NHPPs to make comparisons between the Pre period and the Lockdown period.

Fig. 2-A also shows the time series of the secondary aerosol enhancement, using the ratio of PM$_{2.5}$ major secondary components (including sulfate, nitrate, ammonium and organic matter hereafter SNAO) to the elemental carbon (EC), to eliminate the impacts of the mixing-layer height on pollutant concentrations (18). The ratios of SNAO to EC were stable between the Pre period and Transition period, while a significant rise was found in the Lockdown period under either the HPP or NHPP, with an average increase of 51.8% compared with that in the Pre period. We further investigated the changes in the diurnal variations of SNAO/EC between the Pre period and Lockdown period (Fig. 2-B). The enhancement of secondary aerosols during the COVID-19 outbreak was evident during the entire day with a peak level in the early morning (9:00-10:00 a.m.). This was especially true for nitrate aerosols, which were presented at a peak level more than twice that observed in the Pre period.

In spite of the increases in SNAO during the Lockdown period, the concentrations of NO$_2$ and SO$_2$, regarded as two major gas-phase precursors of secondary PM$_{2.5}$ (nitrate and sulfate), declined by 29.7% and 34.6% on average (Fig. 2-C). Focusing on the differences between NHPPs, the relative reduction of NO$_2$ concentrations reached 57.9% during the Lockdown period compared to the Pre period. This was consistent with the relative reduction of estimated primary emissions (69.3%), indicating that the large emission reduction of local anthropogenic sources actually decreased the PM$_{2.5}$ precursor concentrations during the COVID-19 outbreak.

Ozone is one of the most important oxidants in tropospheric chemistry. As shown in Fig. 2-C, the observed surface ozone increased up to 263.3% between HPPs during the Lockdown compared to the Pre period, with a period-averaged enhancement of 62.0%. In addition, we investigated changes in the nitrate radical (NO$_3$), the primary oxidant for nighttime secondary aerosol formation (19-21). The change in diurnal variations of O$_3$, NO$_3$ radical and NO$_2$ concentrations were provided in SI Appendix, Fig. S7. Compared to the Pre period, NO$_3$ radical concentrations also increased especially at night during the Lockdown period. These changes during the COVID-19 outbreak and the subsequent lockdown indicate that the increased concentrations of oxidants facilitated the chemical formation of secondary fine particles in spite of the significantly reduced gaseous precursors, particularly resulting in a fast nitrate growth during the nighttime. The significant enhancement of the oxidizing capacity was responsible for parts of the rapid growth of secondary aerosols during the Lockdown period.

Our results using the WRF-CMAQ modeling system (BASE scenario for the real-time simulation, see Methods for more details) also revealed a significant increase of oxidant concentrations in most areas of Beijing during the COVID-19 outbreak (SI Appendix, Fig. S8). The O$_3$ variation from the Pre to Lockdown period was investigated using data in the NHPPs in order to reduce the disturbance of regional transport on air quality (Fig. 2-D). The O$_3$ concentration increased by up to 16 parts per billion by volume (ppbv or nmol/mol) covering the northwest to the southeast of Beijing. The estimated vehicular NO$_x$ emission also showed an obvious decrease in the same areas, including urban areas within the 6$_{th}$ ring road and major freight corridors with massive HDTs (Fig. 2-E). In addition, the surface ozone in southern downwind areas also increased most likely due to the prevailing north wind in NHPPs carrying high O$_3$ and precursor gases concentrations from urban areas. As the O$_3$ formation was VOC-limited in urban areas, the relatively larger emission reduction of NO$_x$ compared to VOCs raised the VOC/NO$_x$ ratio (Fig. 1-C), which consequently led to an increase in O$_3$ concentrations (22-24). By contrast, a relatively small emission reduction led to little changes or declines in O$_3$ in western and northern rural areas, where O$_3$ formation is in the transition or NO$_x$-limited regimes (22-24). When focused on emissions within urban areas from different sources, we found that on-road NO$_x$ emission reductions due to the lockdown were responsible for over half (53.4%) of total NO$_x$ emission decrease, while the ratio for VOCs was only 17.8%. Our results indicate that such an imbalance in emission reduction of NO$_x$ and VOC from vehicles was probably responsible for the enhancement of local atmospheric oxidizing capacity, further facilitating the chemical formation of secondary aerosols during the COVID-19 lockdown.

**Impacts of vehicular emission reduction on air pollution during lockdown**

A hypothetical scenario (S1) was set up in which the on-road emissions during the Lockdown period were assumed to be as usual in the Pre period, and other emissions and meteorological conditions were the same as BASE scenario. The differences between BASE and S1 just reflected the impacts from vehicular variations. Fig. 3 shows the changes in spatial distribution of the oxidants and $PM_{2.5}$ concentrations between BASE and S1 from WRF-CMAQ modeling results. A significant enhancement in $O_3$ concentration of up to 11 ppbv was seen in urban areas within the 6th ring road and southern areas, induced by the larger vehicular $NO_x$ emission reduction compared to VOCs. Compared to Fig. 2-D, which shows the $O_3$ variation from the Pre to the Lockdown period, the $O_3$ enhancements only caused by vehicle emissions are similar (Fig. 3-A). It indicates that local traffic emission reduction is the main driving force for the enhanced oxidization capacity in urban areas of Beijing. Compared to changes in $O_3$ concentrations, the increase of $NO_3$ radical was relatively small at the surface level because of the sharp decline of ambient $NO_2$ concentrations (Fig. 3-B), while a more obvious enhancement was found in the upper air within urban areas (approximately 46m above the ground) due to relatively weak NO-titration effects (SI Appendix, Fig. S9). The enhanced oxidants facilitated the formation of secondary organic matter (SOM) during the day and nitrate aerosols in nighttime (SI Appendix, Fig. S10). However, the increased secondary aerosol formation was small, only with a rise by up to 1.6 μg/m3 (Fig. 3-C), probably because of the inadequate precursors particularly during the Lockdown period. While the increased secondary aerosols were still enough to offset the benefit of vehicular reductions in primary emissions, leading to a modest increase of total $PM_{2.5}$ concentrations by up to 1.4 μg/m3 (Fig. 3-D).

Our simulations show that the spatial variations of atmospheric oxidation induced the opposite changes in $PM_{2.5}$ formation in rural (outside the 6th ring road, except for the southeast) compared to urban areas. The enhancement of $O_3$ in rural areas was relatively small, because 80.7%-82.7% of the vehicle emission reductions in Beijing were concentrated in urban areas and the ozone formation regime changed from VOC-limited to $NO_x$-limited going from urban to rural areas. In addition, the reduction of gas-phase precursors restricted the production of secondary particles in rural, resulting in the decrease of SNAO and $PM_{2.5}$ concentrations. Such contrasting impacts between urban and rural areas on both SNAO and $PM_{2.5}$ were more distinct during the HPP compared to those in NHPP, mainly due to unfavorable meteorological conditions for air pollutant dispersion during the HPP (25, 26).

As a conclusion here, the imbalance in emission abatement of $NO_x$ and VOC from vehicles was an important cause for the rise of local atmospheric oxidizing capacity, resulting in a modest enhancement of secondary aerosols and $PM_{2.5}$ concentration. However, the slightly increased $PM_{2.5}$ induced by the giant vehicular emission reduction could not explain the significant growth of secondary aerosols in Beijing during the COVID-19 Lockdown period.

**Why large-scale traffic emission change cannot explain the haze pollution during lockdown?**

The WRF-CMAQ modeling system combined with the Integrated Source Apportionment Model (ISAM) were further applied to trace the contributions of emissions from local sources (on-road vehicles, industry, domestic and others) and regional sources outside Beijing to $PM_{2.5}$ and SNA (the sum of sulfate, nitrate, ammonium) concentrations individually (Fig. 4). Our results show that during the Pre period, the local sources only contributed 19.1% and 30.1% of $PM_{2.5}$ concentrations in winter HPP and NHPP, respectively, which was much less than the results in previous researches with target years of 2012-2014 (44.0%-69.5%) (27-30). This was mainly owning to a more significant reduction of local emissions in Beijing compared to those of surrounding areas in recent years, particularly for the notable decline in local power and heating, industry, and residential sectors from 2013 to 2017 (7). The primary $PM_{2.5}$ from residential sources including both cooking and heating was still the largest contributor among the local sources (hourly averaged 12.0%), followed by "other" local sources (7.5%), in which $NH_3$ emissions forming particulate ammonium were the most important. However, little influence of local vehicles and industry was identified with contributions both less than 3%. As sulfate and nitrate aerosols were more easily transported over long-distance compared with EC, the contribution of regional sources to SNA concentration reached 89.1% and 84.9% during the HPP and NHPP during the Pre period, respectively.

During the Lockdown period, the contribution of regional transport to ambient $PM_{2.5}$ increased to 86.8% and 88.5% during the HPP and NHPP, respectively, and it was responsible for more than 90% of SNA concentrations, since more emissions from industry and residential sectors were reduced in Beijing compared to its neighboring regions during the COVID-19 Lockdown (SI Appendix, Table S1). Our modeling results showed that the enhancement of atmospheric oxidizing capacity during the Lockdown period was recognized in not only Beijing, but also most areas of the Beijing-Tianjin-Hebei (BTH) and its surrounding regions (SI Appendix, Fig. S11), consistent with recent research (31). Differing from "significant increased $O_3$ but modest enhanced secondary aerosols" in Beijing, the increased oxidants in the case of relatively abundant gaseous precursors outside Beijing facilitated more chemical formation of secondary organic and inorganic aerosols, which were transferred into Beijing and significantly increased the local secondary $PM_{2.5}$ concentrations. The contributions of almost all local sources to $PM_{2.5}$ and SNA concentrations decreased sharply during the Lockdown period, while the contribution of "other" local sources still remained at a normal level due to the important role of its $NH_3$ emissions in the chemical growth of secondary inorganic aerosols. As a result, although traffic activities and emissions in Beijing were significantly reduced during the pandemic, it still could not turn over the aggravated haze pollution, due to the complicated nonlinear response of the fine particle and air pollutant emissions as well as the dominant impacts of regional sources.

**Discussion on traffic emission control perspectives**

Beijing "pandemic haze" is a challenging case for source-receptor relationships (Fig. 5). Large emissions from industry in the BTH region (including boilers, cement, steel production and other industrial processes), as well as the further increased emissions in winter for heating (32, 33), provide sufficient precursors to form the secondary aerosol in the case of enhanced oxidants, which transported into Beijing, resulting in an aggregated haze pollution. Even during the Lockdown period, most of the heavy industries in Hebei province were still in continuous operation with little emission reductions (34). On the other hand, since Beijing is already in the "low concentration pool" of BTH region, a significant enhanced local atmospheric oxidizing capacity caused by the imbalanced vehicular emission reduction of $NO_x$ and VOC, only leads to a modest increase of secondary aerosols and $PM_{2.5}$ concentration due to the inadequate precursors. All these make the relationship of vehicle emissions and air quality in Beijing different from that in other megacities (35).

The public were disappointed by the "pandemic haze" because of an expectation that previous efforts on controlling traffic to avoid the $PM_{2.5}$ pollution would mean that during the lockdown there would be no pollution problems. Our modelling results indicate that the local traffic activity had little impacts on the winter haze events in Beijing. In addition, based on a series of sensitivity runs modifying the local on-road emissions in winter normal days (Pre period), even without vehicle emissions, the ambient $PM_{2.5}$ concentration could only be reduced by 1.2 μg/m$_3$ on average in urban and southern rural areas of Beijing (SI Appendix, Fig. S12). Although vehicles accounted for 65.6% of local $NO_x$ emissions, the concentrations of oxidants and fine particles would be consistently enhanced, with the reduction ratio of the vehicular $NO_x$ emissions rising from 0% to 100%, even when VOC emissions from vehicles were reduced by 100%. On the one hand, the abundant nitrate precursors in neighboring regions suppresses the effectiveness of local $NO_x$ emission control. On the other hand, reducing $NO_x$ emission favors the enhanced atmospheric oxidation ability to form more secondary particles, since many urban areas in China are prevailing under the VOC-limited condition (36). Although reducing vehicular VOC and primary PM emissions were both positive in decreasing the $PM_{2.5}$ levels, unfortunately, traffic control usually leads to greater $NO_x$ reduction than VOCs, which goes to explain why the annual reduction of vehicular emissions has resulted in reductions in VOCs that are only half (in percentage terms) the $NO_x$ reductions over the past years (37). All these points above explain why traffic control cannot mitigate the winter haze pollution in Beijing currently, a point which also needs to be better explained to the public.

In past years, the gradually strengthened vehicle emission controls have successfully contributed to the PM$_{2.5}$ decrease in China (38). In addition, past experiences from developed countries indicates that emission control on the continuously growing motor vehicle fleet is efficient and ultimately cost-effective to relieve the air pollution in a megacity (9, 10). The problem for now is the imbalanced control among different source regions and air pollutants. Compared to its surrounding regions, local sources in Beijing reduced faster in the past years, which lead to the share of regional transport for the air pollution in Beijing increasing to 76.1% on average in the winter. As for the differences in local species controls in Beijing, the reduction rate of NO$_x$ and VOCs emissions used to be approximately 1:1, while the ratio is still large enough to increase the atmospheric oxidizing capacity under the strong VOC-limited condition in winter, which is proved by the enhanced oxidants during the COVID-19 lockdown (7, 39). In addition, NH$_3$ emissions, which are crucial for the formation of secondary inorganic aerosols, were still not effectively controlled, and the increased NH$_3$ concentrations over past years weakened the benefits of the reduction in nitrate from NO$_x$ emission control in East China (40). Therefore, the key is not judging whether traffic emission control is necessary, but accelerating VOC and NH$_3$ emissions reductions and synchronously controlling regional sources to release the benefits on local traffic emission control.

Targeting any of long-term air quality, climate change or street-level personal exposure, means any measures reducing vehicle emissions are going to be beneficial. To achieve PM$_{2.5}$ concentration reductions in the short term, VOCs and primary PM$_{2.5}$ should be jointly treated as the priority pollutants to control, or NO$_x$ emissions should be substantially reduced by the combination with other sources beyond vehicles so as to reach the non-linear tipping-point between changes of NO$_x$ reduction and oxidants concentration. For the first option, the new emission standard for LDV (China 6), which will be implemented in July 2020, is expected to dramatically reduce the VOC emission from evaporations and will be effective to improve the air quality (41).

The challenge is to maintain popular support for mitigation policies such as reductions in traffic flow or restrictions in vehicle type which themselves lead to significant air quality improvements (PM$_{2.5}$, NO$_2$) but which are not directly visible to the general population in the face of the highly visible haze events

## Materials and Methods

In this work, we built an integrated analysis framework to seek for the role of vehicle emissions in winter haze pollution in Beijing around the COVID-19 outbreak and Spring Festival (SI Appendix, Fig. S13). First, we developed a SLOVE model to estimate the hourly traffic emissions in urban areas of Beijing. This model consists of two dynamic databases, including a) hourly road speed and b) the observed meteorological condition, and three static local traffic information databases, including a) fleet composition, b) road basic information and c) vehicle emission factors. The real-time traffic condition data through the Application Programming Interface (API) to AMap was obtained to calculate the traffic flow and emissions. Detailed descriptions about this model were discussed in the SI Appendix, section S1. In addition, the emissions from HDTs were evaluated based on a more accurate TrackATruck model driven by big data (trajectory signals of each HDT from the BeiDou Navigation Satellite System), with advantages of considering individual truck differences (42).

Besides the on-road emissions calculated in this study, emissions from other sectors were assembled from several recent studies to improve the precision of emission inventory, including urban anthropogenic emissions for Beijing, Tianjin and its surrounding 26 major cities in northern China from an air pollution prevention plan proposed by the government (referred to "2+26" plan), shipping emissions from our previous research (43), other anthropogenic emissions in China from Multi-resolution Emission Inventory for China (MEIC) model (38) and others listed in SI Appendix, Table S2. The emission changes of other anthropogenic sources during the transition and lockdown period were calculated respectively, based on the changes of related industrial and residential activities (SI Appendix, sections S2).

The Weather Research and Forecasting Model–Community Multiscale Air Quality Model (WRF-CMAQ) was applied to simulate the air quality in Beijing from January $10_{th}$ to February $15_{th}$ in 2020 (44, 45). The modeling system drew on the 4-nested run with a grid resolution at 1.33 km of the innermost domain, where Urban Canopy Model (UCM) with updated land use data and Urban Canopy Parameters (UCPs) was applied to improve the prediction of meteorological field (46). To reproduce the polluted days, the heterogeneous reactions of $SO_2$ were incorporate into the CMAQ model to enhance the sulfate formation at a high relative humidity (47, 48). In addition, the CMAQ model with the ISAM was used to determine the source contribution to the $PM_{2.5}$ and its species concentrations before and during the COVID-19 outbreak. Detailed model configurations were described in SI Appendix, section S3.

The ground meteorological observations were obtained from the National Climate Data Center (NCDC, ftp://ftp.ncdc.noaa.gov/pub/data/noaa/) integrated surface database. The hourly air quality data in Beijing, including $PM_{2.5}$ and gaseous air pollutants were from the Beijing Municipal Environmental Monitoring Center (BJMEMC). The $PM_{2.5}$ component observations were collected from the National Research Program for Key Issues in Air Pollution Control. We evaluated the simulated $NO_2$, $O_3$, and $PM_{2.5}$ concentrations against ground-based observations. The model well captured the variations of the air quality with correlation coefficients higher than 0.5 for all species, which was in line with other recent modeling studies in Beijing (49, 50). The model performances in predicting major $PM_{2.5}$ chemical components were also acceptable with the mean biases (MB) ranging from 0.3 μg/m$_3$ to 4.3 μg/m$_3$.

To quantify the influence of vehicle emission reduction on air quality before and during the COVID-19 outbreak, we set up a series of scenarios in the WRF-CMAQ modeling system with different turbulences in on-road emissions while other configurations remained the same as the BASE scenario. The BASE scenario simulated the air quality with real emissions, while in S1 scenario vehicle emissions during the Lockdown period were assumed to be as usual in the Pre period. In S2-S6 scenarios, the assumed relative reduction of vehicle emissions changed from 0% to 100% in winter normal days (Pre period).

Our study was subject to a few uncertainties and limitations. Besides uncertainties of emission inventories and WRF-CMAQ models (discussed in SI Appendix, section S4 and S5), the SOM were not considered in the source apportionment due to the limitation of the existing ISAM model. We probably underestimated the contribution of local vehicle emissions to PM$_{2.5}$ concentrations since (a) aromatics from gasoline vehicle exhausts is a critical determinant of urban secondary organic aerosol formation and (b) synergetic oxidation of vehicular exhaust leads to efficient formation of ultrafine particles (UFPs) under urban conditions (51, 52). Moreover, the on-road emissions are instantly diluted throughout the coarse grid cells of the chemical transport model in which the emissions occur. In a finer spatial scale, however, the vehicle emission would significantly affect the human exposure to air pollution due to its close proximity to human activities at a low emission height, and thus result in a serious health burden (53). Quantification of vehicular contribution to human health risk in Beijing at a neighbourhood-scale based on a source-receptor model with a higher spatial resolution is necessary and suggested for future investigation.

## Acknowledgments

This work was supported by the National Natural Science Foundation of China (No. 41822505, 42061130213), the Royal Society of UK through Newton Advanced Fellowship (NAF\R1\201166), the Tsinghua-Toyota General Research Center, the Beijing Nova Program (Z181100006218077), the Tsinghua University Initiative Scientific Research Program and the Foshan-Tsinghua Innovation Special Fund (FTISF-2019THFS0402), National Research Program for Key Issues in Air Pollution Control (DQGG0201&0207).

## Author Contributions

H.L and Z.L. designed this research and performed analysis. Z.L. and X.W wrote the paper. All authors took contributions to discussing and improving this research.

# Figures and Tables

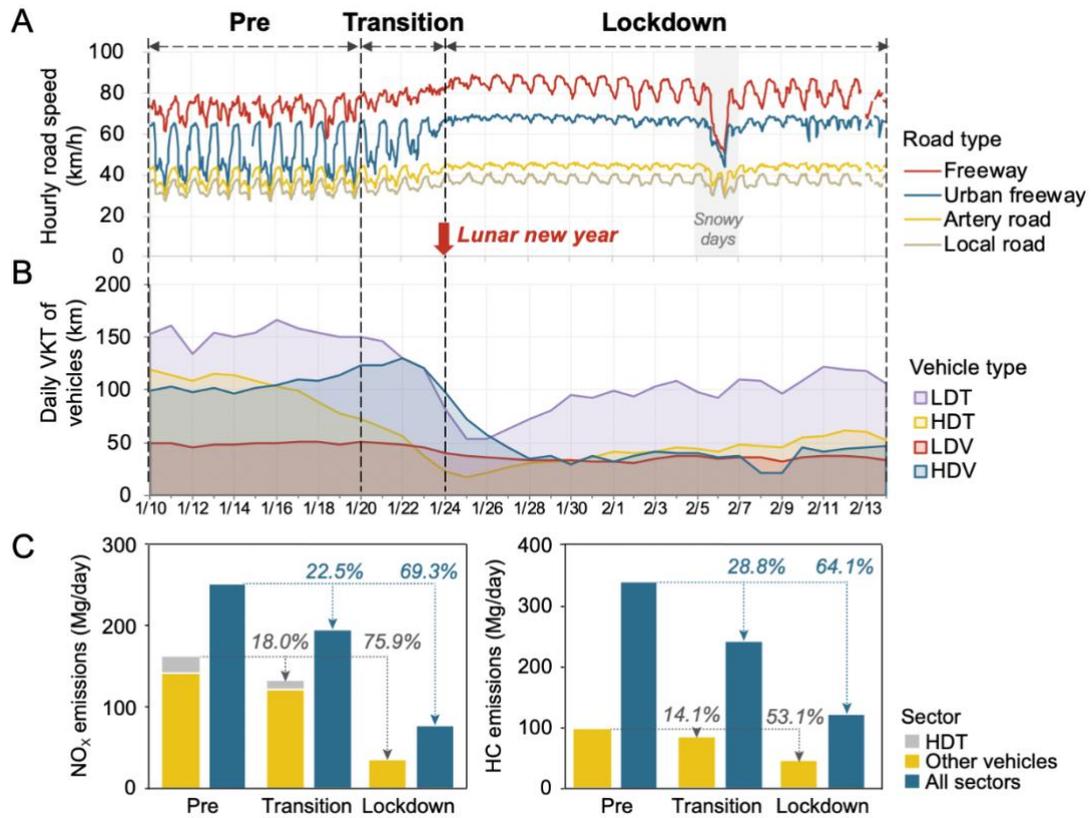

**Figure 1.** Traffic activities and emissions in Beijing around the COVID-19 outbreak. (A) Hourly road speed within the 6th ring road of Beijing. (B) Daily VKT of different vehicle types. (C) Daily emissions of $NO_x$ and HC from vehicles and all sources during different periods. The percentages in (C) stand for the relative changes in vehicle emissions compared to the Pre period.

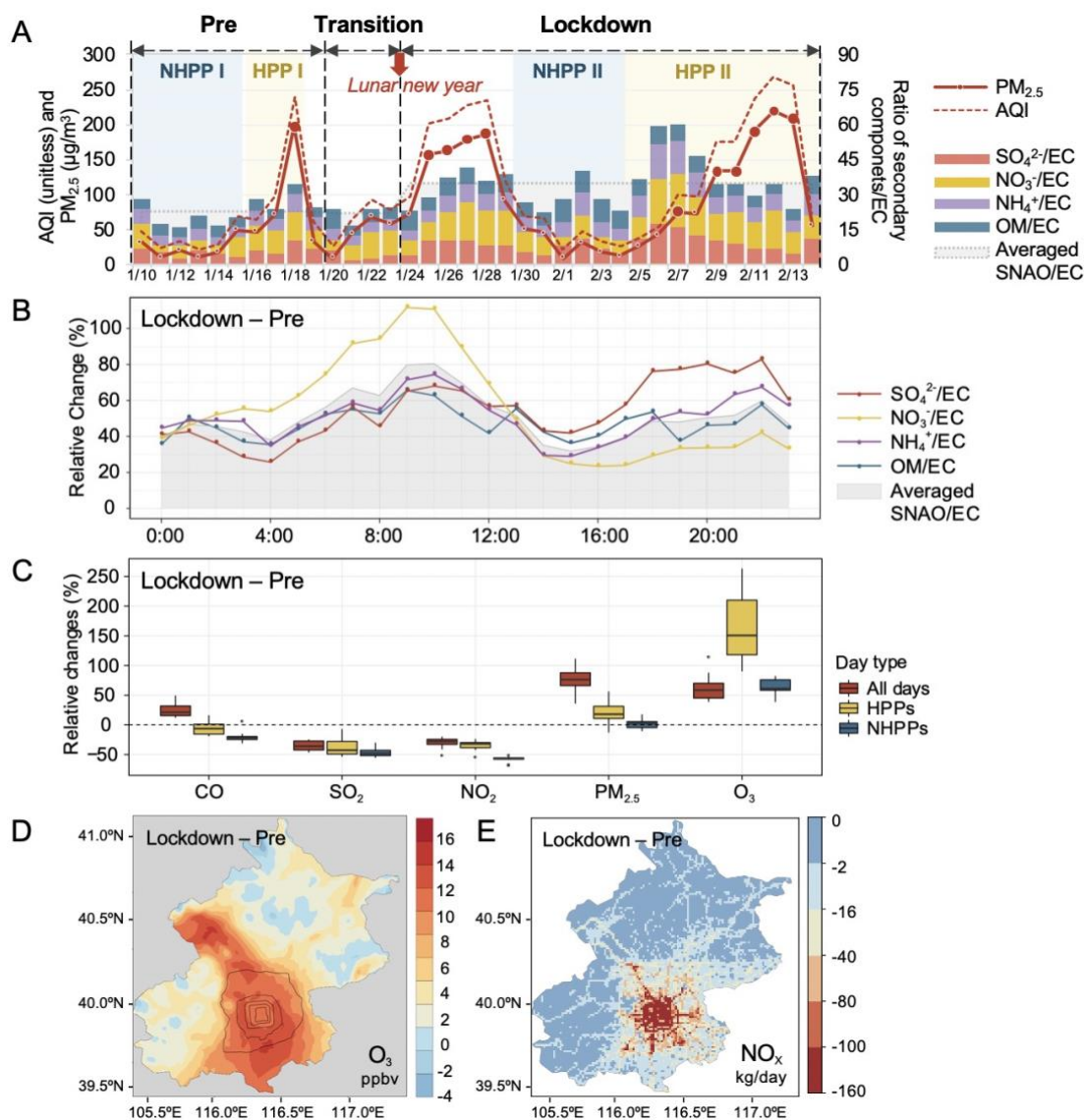

**Figure 2.** Changes from the Pre period to the COVID-19 Lockdown period. (A) Observed AQI, $PM_{2.5}$ concentrations, and the ratios of $PM_{2.5}$ secondary components/element carbon ($EC$), including organic matter ($OM$), sulfate ($SO_4^{2-}$), nitrate ($NO_3^-$) and ammonium ($NH_4^+$). (B) Observed diurnal relative changes of the SNAO/EC. (C) Observed relative changes in pollutant concentrations in all monitoring sites of Beijing. (D) Spatial changes of modeled $O_3$ concentrations (E) Spatial changes of calculated vehicular $NO_x$ emission reduction. The larger points in (A) indicated daily average $PM_{2.5}$ concentrations were larger than 75 μg/m3.

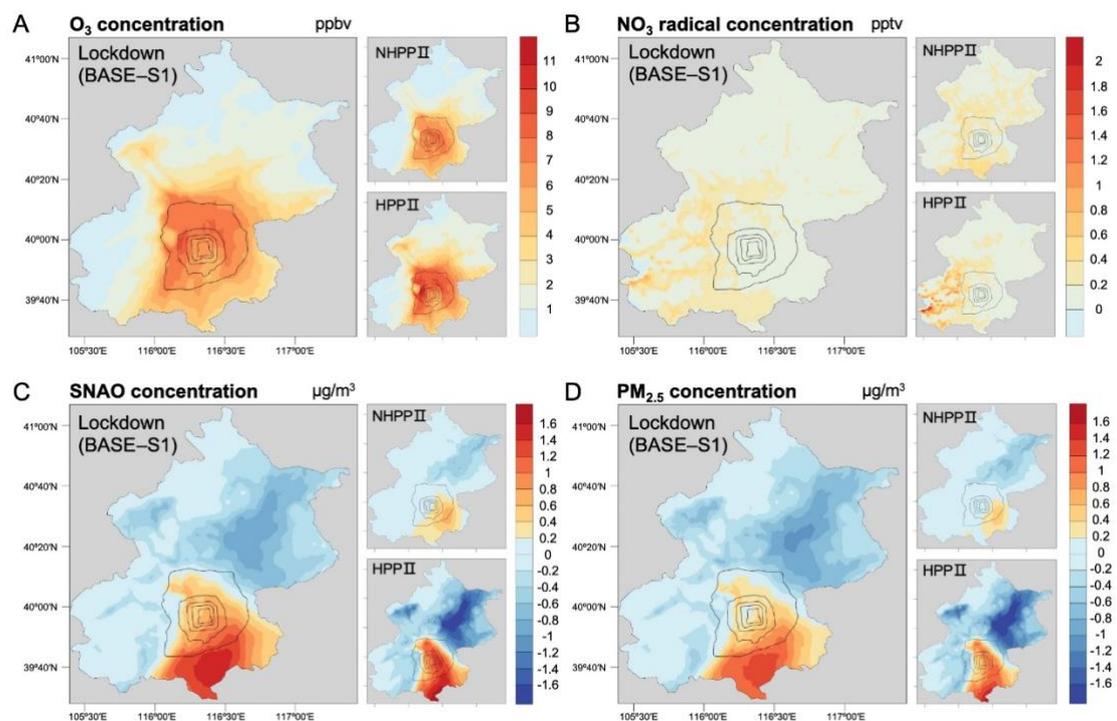

**Figure 3.** Changes in modeling concentrations between the real-time emission scenario (BASE) and the traffic-as-usual emission scenario (S1) during lockdown period (BASE-S1). (A) $O_3$ (B) $NO_3$ radical (C) SNAO (the sum of sulfate, nitrate, ammonium, and secondary organic matter) (D) $PM_{2.5}$. The black lines represent the ring roads of Beijing. Plots are for model surface level.

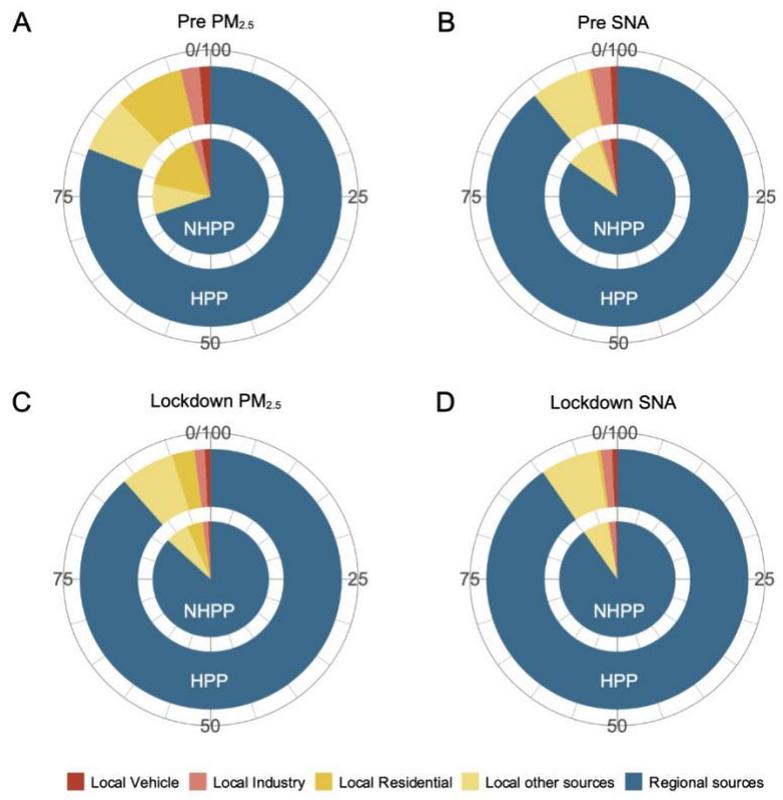

**Figure 4.** The contribution of emissions from local and regional sources on $PM_{2.5}$ and SNA concentrations in Beijing during the Pre period and lockdown period, units in %. The inner and outer pie charts stand for the source contribution during the NHPP and HPP, respectively.

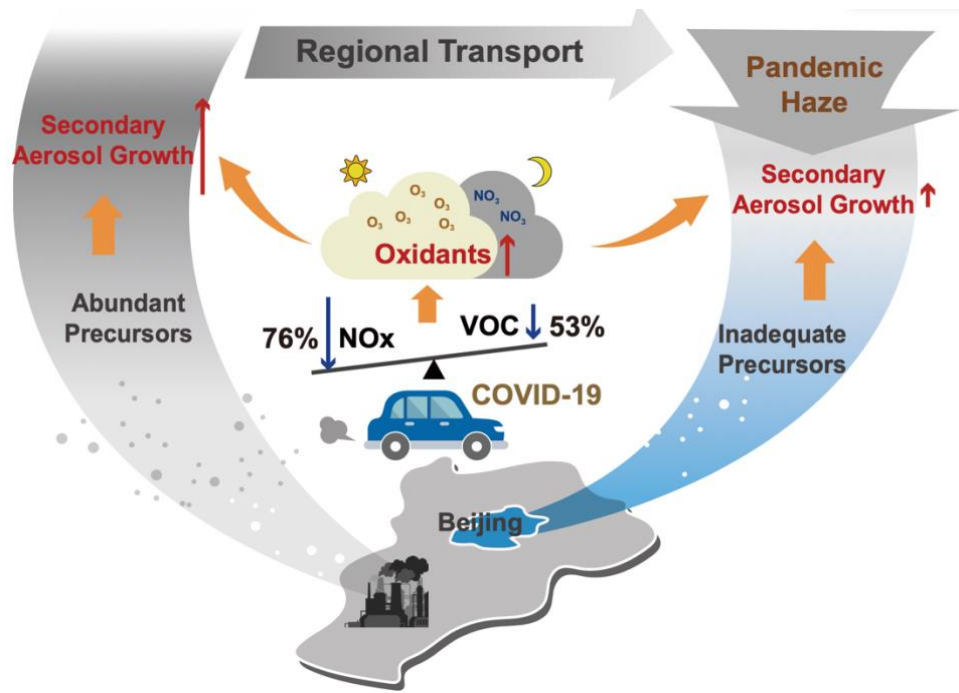

**Figure 5.** Simplified schematic of the role of reduced vehicle emissions in the pandemic haze formation in Beijing.

**Supplementary Results**

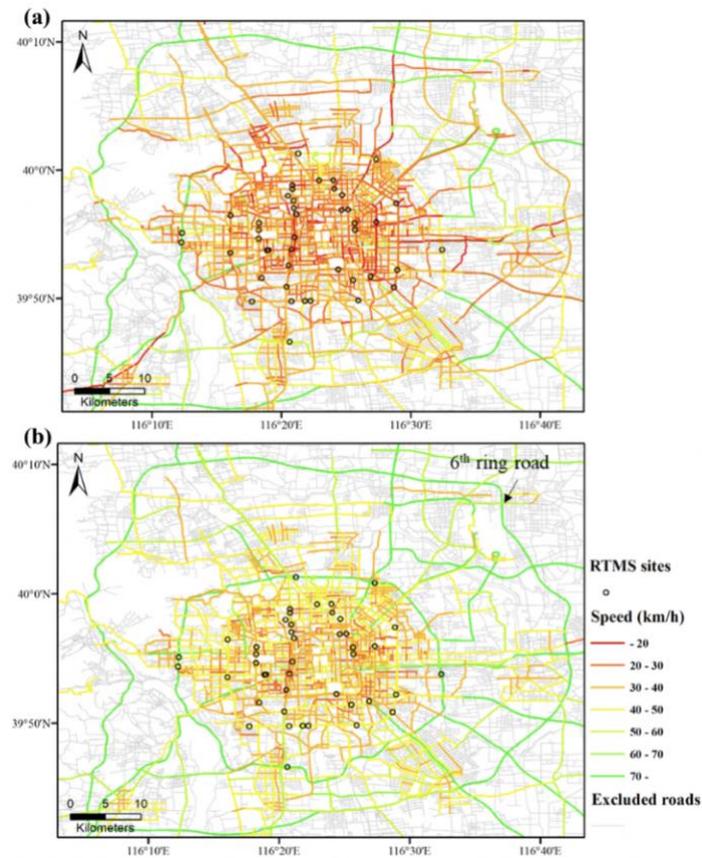

**Fig. S1.** Spatial distribution of traffic speeds extracted from AMap within the 6th ring road of Beijing at morning rush hours (8:00-9:00 a.m.) on (a) January 13th (Monday, Pre period) and (b) February 3rd (Monday, Lockdown period) in 2020. The circles represent the traffic monitors collected by the Beijing Municipal Commission of Transport (BMCT) using the Remote Traffic Microwave Sensor (RTMS). The gray lines represent the non-monitoring roads.

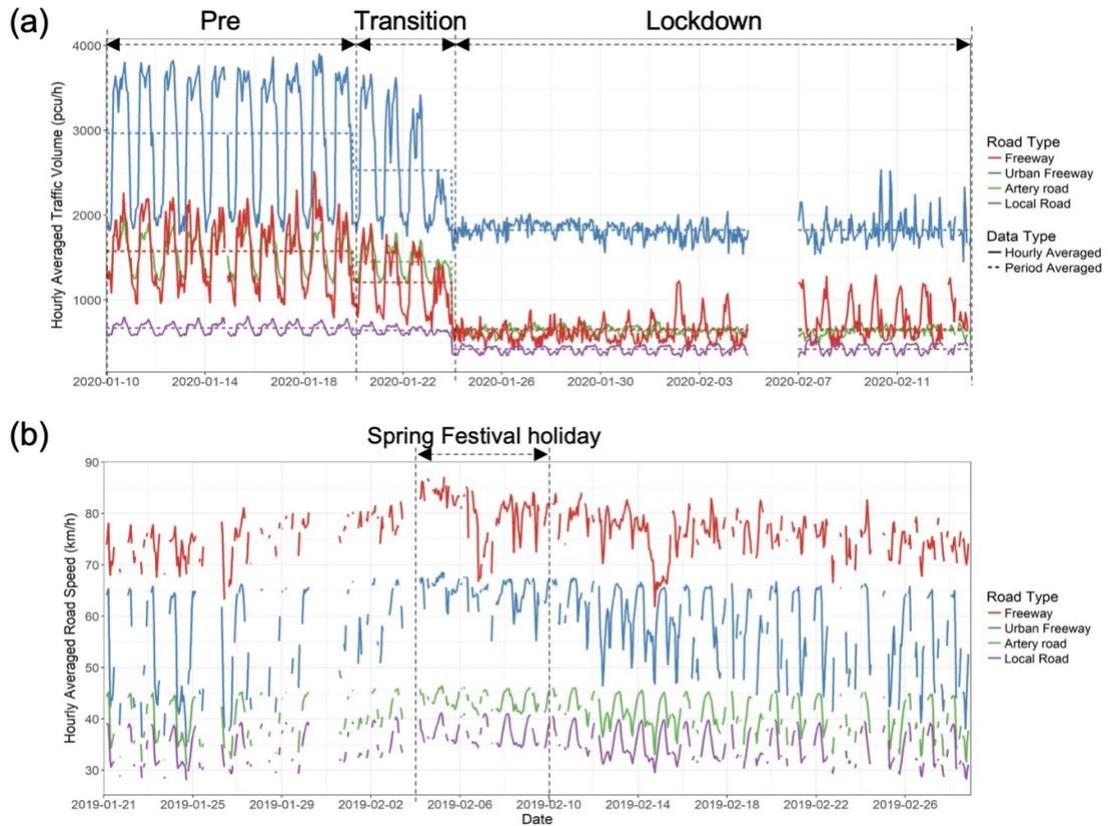

**Fig. S2.** Hourly averaged traffic condition of various roads within the 6th ring road of Beijing. (a) traffic flow in 2020 estimated by the SLOVE model (b) hourly road speed in 2019 extracted from the AMap. Note: Breakpoints on lines represented missing values due to the temporary brokedown of the equipment or network, except for the removal of traffic volumes from 5th Feb. to 6th Feb, when the uncertainties of the SLOVE model were relatively large during snowy days.

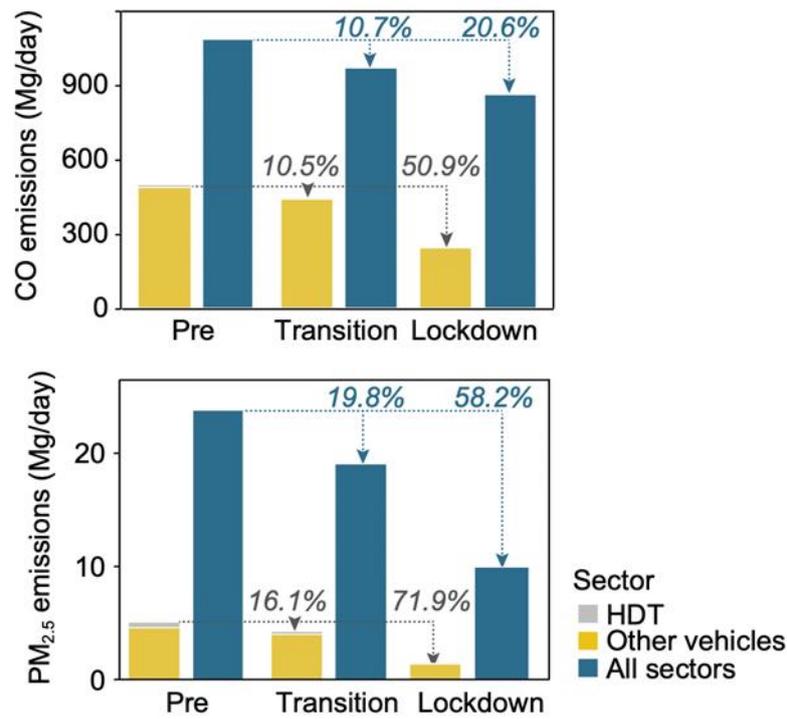

**Fig. S3.** CO and PM$_{2.5}$ vehicle emissions of Beijing in the Pre period, Transition period and COVID-19 Lockdown period. The percentages stand for the relative changes in vehicle emissions compared to Pre period.

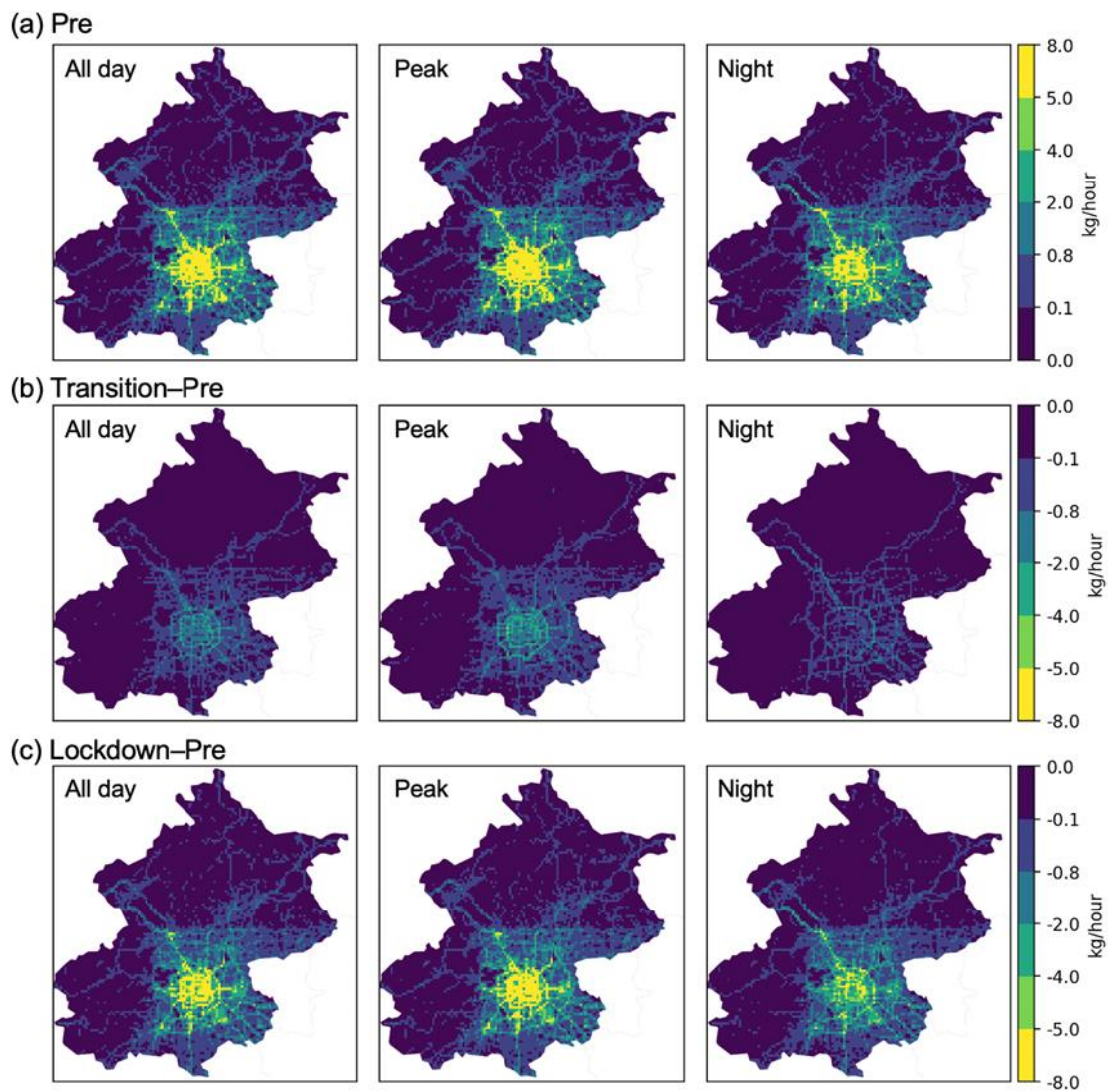

**Fig. S4.** Spatial distributions of hourly averaged on-road emissions of Beijing at 1 km×1 km grids during all day, traffic peak hours (7:00 a.m.-9:00 a.m. and 5:00 p.m.-7:00 p.m.) and night hours (0:00 a.m.-6:00.a.m.) for (a) Pre period, (b) changes between Pre and Transition period, and (c) changes between Pre and Lockdown period.

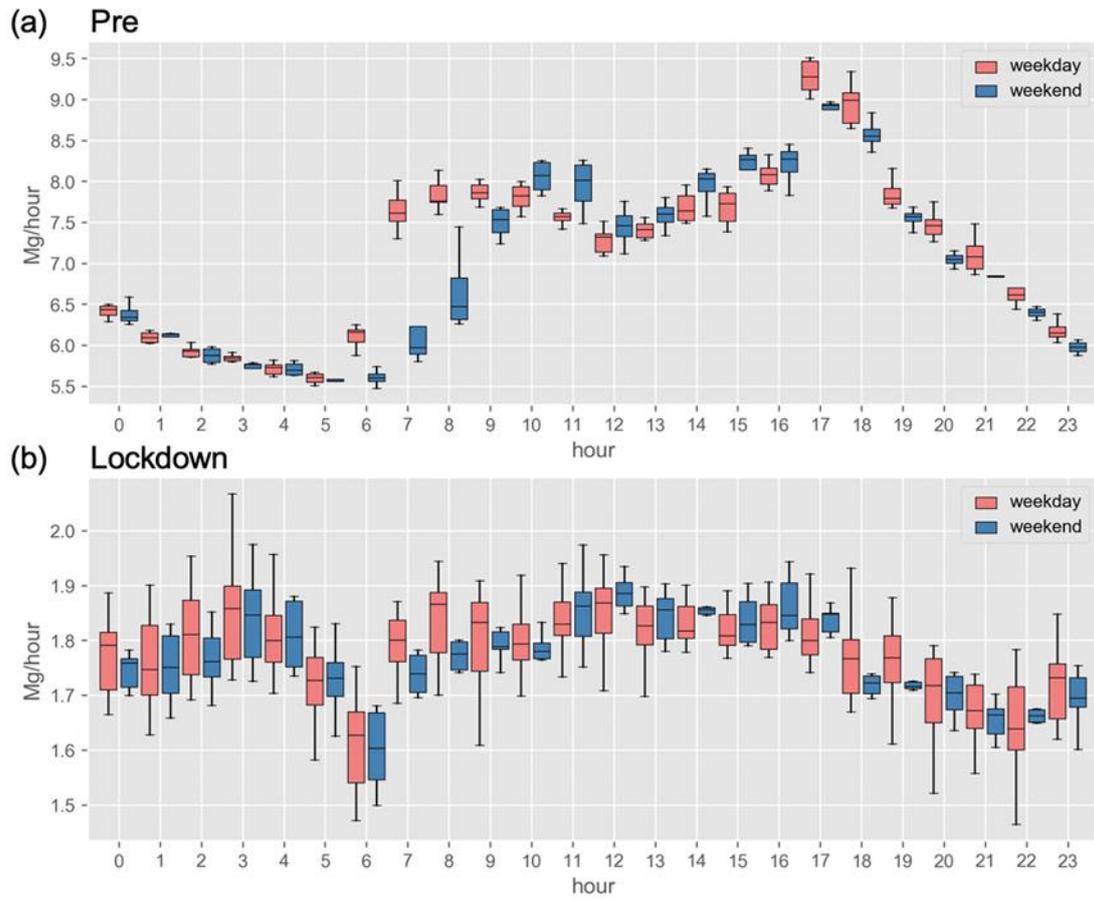

**Fig. S5.** Time series of on-road emissions on weekdays and weekends in Beijing during (a) Pre period and (b) COVID-19 Lockdown period. Samples were defined as hourly traffic emissions for each day during different periods. The interval between the upper and lower cap line represents the 5–95% sample distribution. The box represents the 1/4 to 3/4 quantile sample distribution.

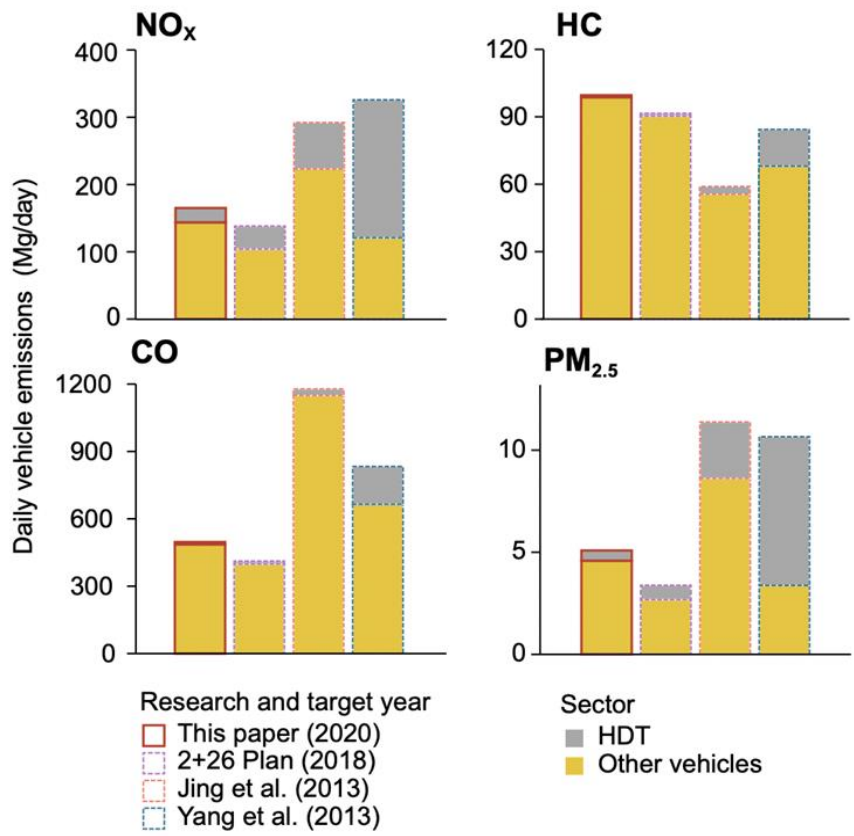

**Fig. S6.** Comparison of vehicle emissions of Beijing during the Pre period in this study with previous studies.

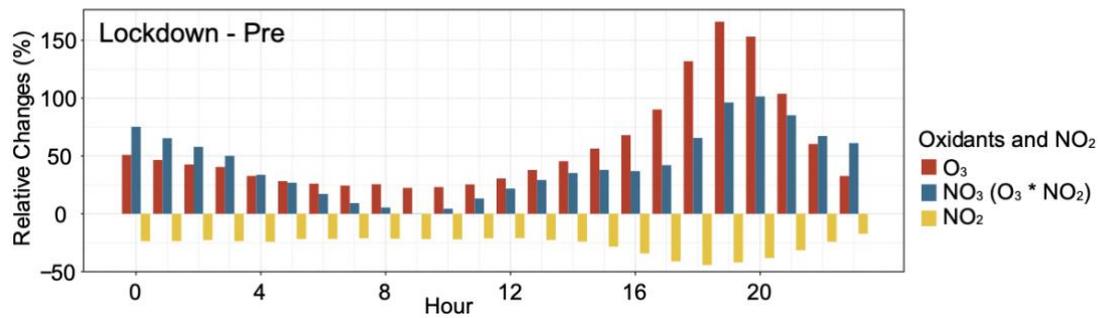

**Fig. S7.** Observed changes of diurnal variations of oxidants and $NO_2$ between Pre and Lockdown period. A proxy $O_3 \times NO_2$ was used to represent changes in the nitrate radical ($NO_3$).

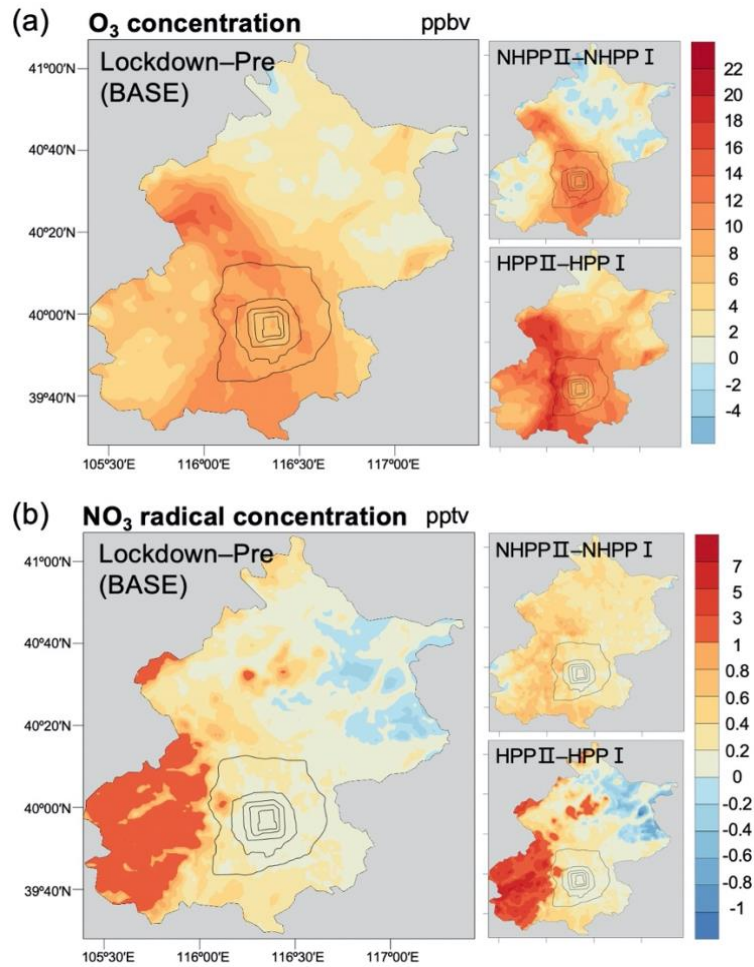

**Fig. S8.** Changes in the modeling oxidants between Pre and Lockdown period in Beijing (BASE scenario). (a) $O_3$ concentration (b) $NO_3$ radical concentration. The black lines represent the ring roads of Beijing.

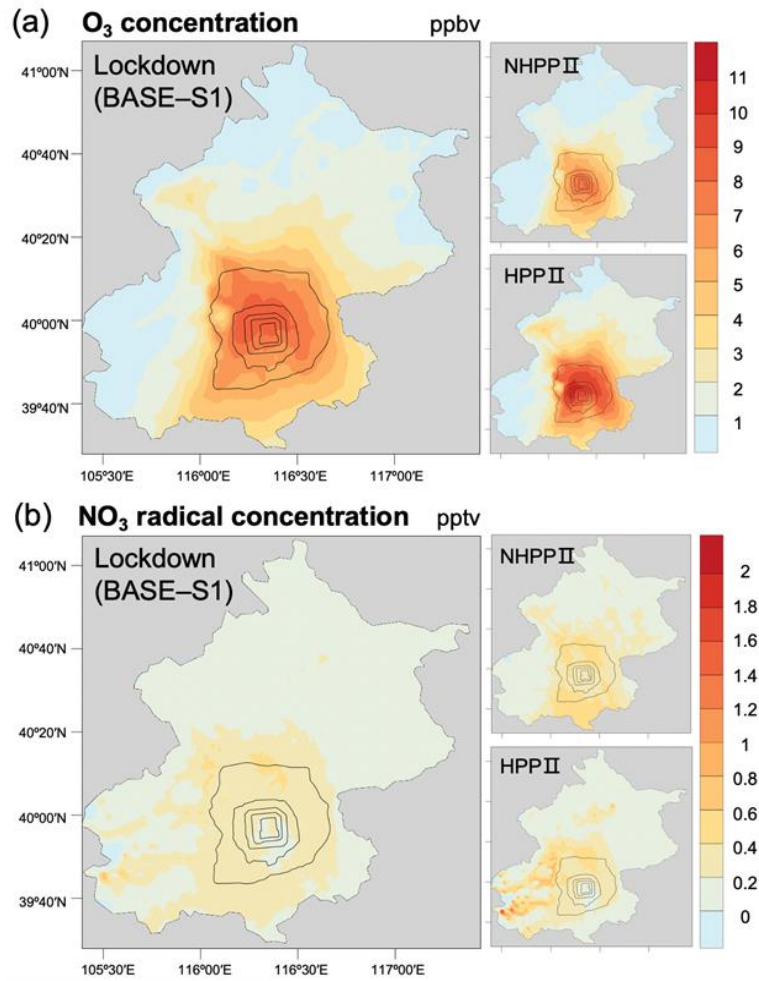

**Fig. S9.** Changes in the modeling oxidants at approximately 46m above the ground between the real-time simulation (BASE) and the traffic-as-usual emission scenario (S1) during the Lockdown period (Base-S1). (a) $O_3$ concentration (b) $NO_3$ radical concentration. The black lines represent the ring roads of Beijing.

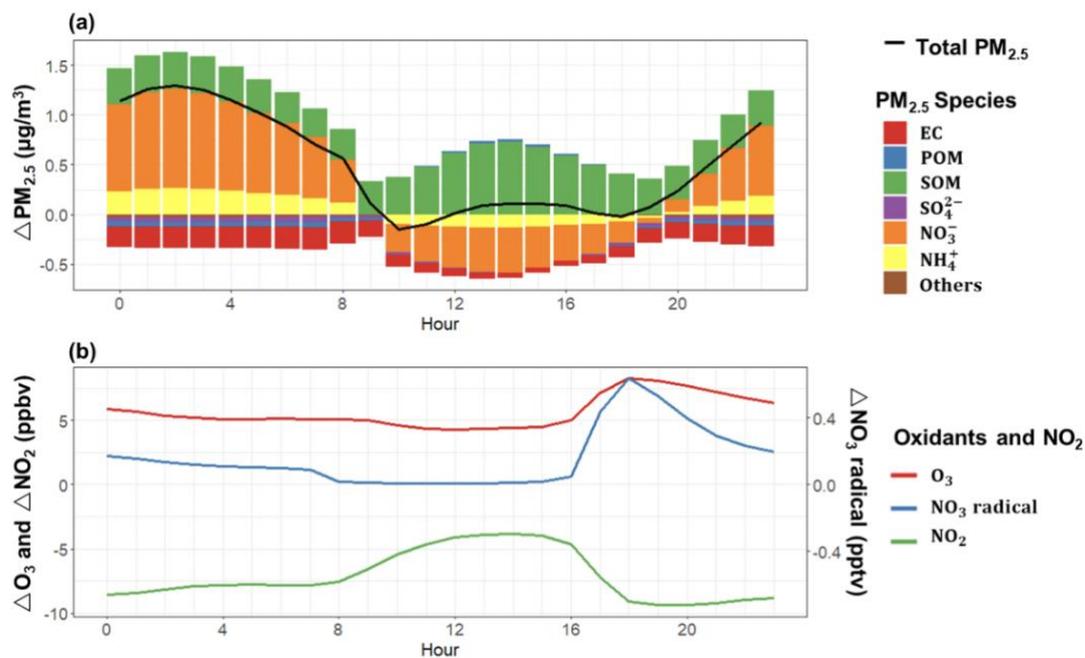

**Fig. S10.** Modeling changes of diurnal variations of PM$_{2.5}$ species, oxidants, and NO$_2$ concentrations between BASE and S1 during the Lockdown period in urban and southern rural areas (39.44°N-40.18°N, 116.09°E-116.71°E) (BASE-S1). (a) PM$_{2.5}$ and its major components (b) Oxidants and NO$_2$. The major PM$_{2.5}$ species included element carbon ($EC$), primary organic matter ($POM$), secondary organic matter ($SOM$), sulfate ($SO_4^{2-}$), nitrate ($NO_3^-$) and ammonium ($NH_4^+$).

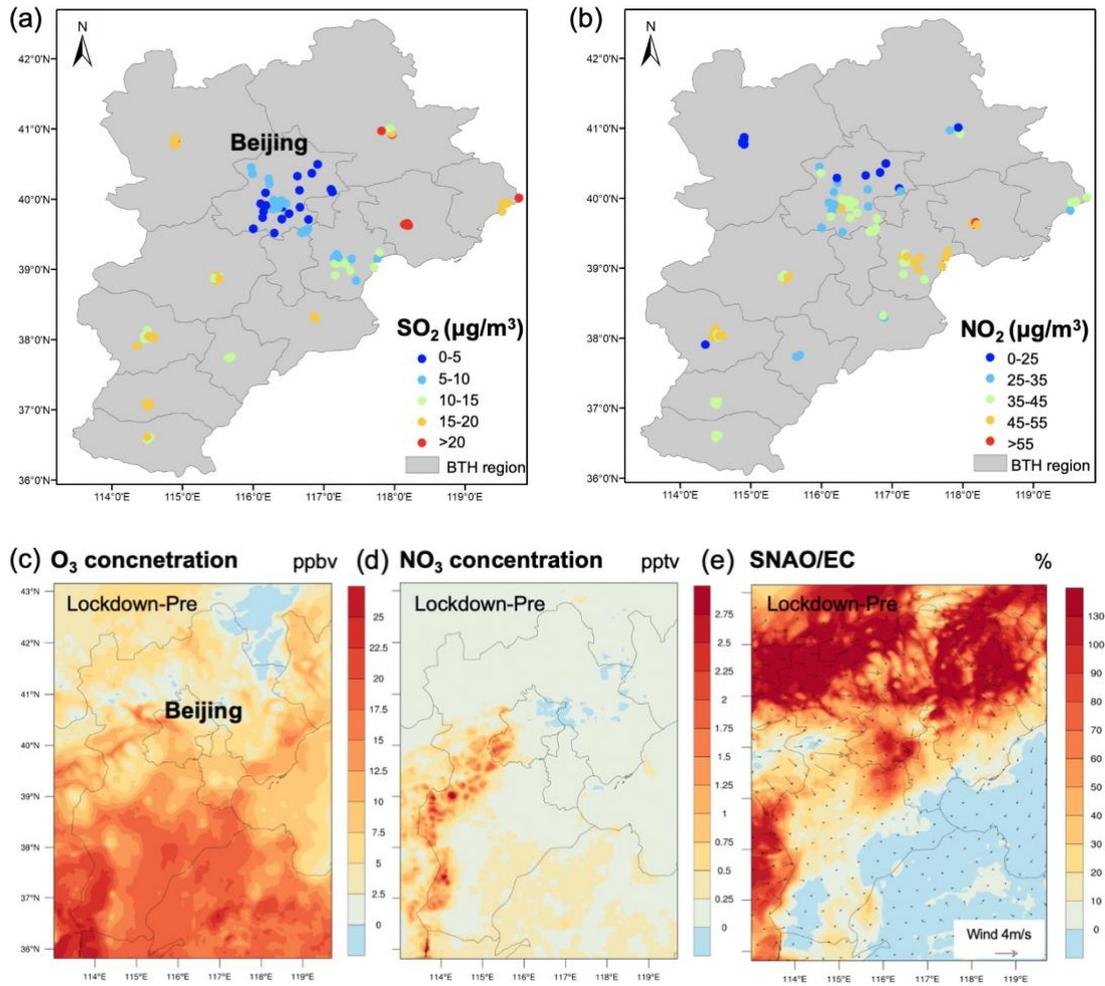

**Fig. S11.** Observed and modeled concentrations of air pollutants in BTH (Beijing-Tianjin-Hebei) region. Observed (a) $NO_2$ concentration and (b) $SO_2$ concentration of monitoring sites for the overall research period. Modeling (c) $O_3$ concentration (d) $NO_3$ radical concentration (e) SNAO/EC changes between Pre and Lockdown period (results from d03 BASE scenario). Arrows in (e) represented the mean surface wind field during Lockdown period from WRF modeling results by vector-averaging.

| Scenarios | Relative reduction of vehicle emissions (%) | | | PM$_{2.5}$ (μg/m$^3$) | O$_3$ (ppbv) | NO$_3$ radical (pptv) |
|---|---|---|---|---|---|---|
| | NOx | VOC | PM$_{2.5}$ | all-day | all-day | nighttime |
| S2-BASE | -76 | -53 | -72 | 0.1 | 3.4 | 0.1 |
| S3-BASE | -100 | -100 | -100 | -1.2 | 5.2 | 0.3 |
| S4-BASE | -100 | -100 | 0 | 0.8 | 5.2 | 0.3 |
| S5-BASE | -50 | -100 | 0 | 0.1 | 1.8 | 0.1 |
| S6-BASE | 0 | -100 | 0 | -0.3 | -0.1 | -0.0 |

**Figure S12.** Averaged changes of PM$_{2.5}$, O$_3$, and NO$_3$ radical concentrations due to the different vehicle emission reduction scenarios (S2-S6) in urban and southern rural areas of Beijing (39.44°N-40.18°N, 116.09°E-116.71°E).

**Table S1.** Relative emission reduction from anthropogenic sources during COVID-19 Transition and Lockdown period compared to the Pre period in Beijing and other regions.

| Sector | | Transition | | | | | | Lockdown | | | | | |
|---|---|---|---|---|---|---|---|---|---|---|---|---|---|
| | | CO | NH$_3$ | NO$_x$ | PM$_{2.5}$ | SO$_2$ | VOC | CO | NH$_3$ | NO$_x$ | PM$_{2.5}$ | SO$_2$ | VOC |
| Power | Beijing | - | - | - | - | - | - | - | - | - | - | - | - |
| | Other | - | - | - | - | - | - | - | - | - | - | - | - |
| Industry | Beijing | 39% | 39% | 39% | 39% | 39% | 39% | 71% | 71% | 71% | 71% | 71% | 71% |
| | Other | - | - | - | - | - | - | 20% | 20% | 20% | 20% | 20% | 20% |
| Residential | Beijing | 18% | 18% | 18% | 18% | 18% | 18% | 35% | 35% | 35% | 64% | 35% | 59% |
| | Other | - | - | - | - | - | - | - | - | - | - | - | - |
| Transportation | Beijing | 9% | 14% | 14% | 13% | 14% | 14% | 50% | 53% | 74% | 70% | 53% | 53% |
| | Other | - | - | - | - | - | - | 50% | 53% | 74% | 70% | 53% | 53% |
| Agriculture | Beijing | - | - | - | - | - | - | - | - | - | - | - | - |
| | Other | - | - | - | - | - | - | - | - | - | - | - | - |
| Open burning | Beijing | - | - | - | - | - | - | - | - | - | - | - | - |
| | Other | - | - | - | - | - | - | - | - | - | - | - | - |
| Ship | Beijing | - | - | - | - | - | - | - | - | - | - | - | - |
| | Other | - | - | - | - | - | - | - | - | - | - | - | - |

# Supplementary Methods
## Section S1 Street-Level On-road Vehicle Emission (SLOVE) model

With the rapid development of intelligent transportation system (ITS) technologies, the floating car data is collected by the digital map providers (such as Amap, Baidu, and Didi in China) from the GPS of commercial taxies and the signals received from the map user's mobile devices. Therefore, this kind of dynamic traffic big data can represent the real-time on-road conditions, which has been successfully used to estimate vehicle emissions in megacities with high data coverage due to a large number of users (1, 2). Based on the local traffic characteristics of Beijing, we developed a Street-Level On-road Vehicle Emission (SLOVE) model to estimate the hourly traffic emissions. First, to obtain the hourly spatial mean speed of urban roads, we accessed the real-time traffic data through the Application Programming Interface (API) to the Amap (www.amap.com) 12 times per hour to improve the temporal coverage of the data. Based on the traffic function and capacity from high to low, the roads were classified into four types, including freeway, urban freeway, artery road, and local road. The data collected on Monday morning rush hours in Pre period and the COVID-19 Lockdown period were taken as an example to show the spatial distribution of traffic speeds (Fig. S1). We found that the improvement of traffic conditions after the COVID-19 outbreak was obvious due to the epidemic prevention, indicating that our data was reliable to reflect the real-world road information.

The road traffic flow was then estimated based on local speed-volume models. Previous studies indicated that the single-regime models from traffic engineering could be applied to describe the relationship between traffic speed and flow in urban roads of Chinese cities (1-4). Although there are many other advanced models, most of them are too complicated to use in a megacity like Beijing. In this study, we assumed that the same speed-volume model could be applied to roads with the same road type because of similar traffic capacity. The speed-volume model in the urban freeway was established based on the observations from 45 sites with more than 200,000 samples collected by the Beijing Municipal Commission of Transport (BMCT) using the Remote Traffic Microwave Sensor (RTMS) (Fig. S1). The historical monitoring data included the hourly traffic speed and flow for each day, covering not only four major ring roads but also some non-ring roads, collected from February 2018 to October 2018. The fitting performances among the Underwood model, Greenshields model, and Van Aerde model were compared (5-7). The results showed that the Van Aerde model was the best one to reproduce the traffic flow for all selected roads with Root Mean Square Error (RMSE) in the range of 144-325 pcu·(hour·lane)$_{-1}$ (Fig. S14). For other road types lack of observations, the fitted Underwood models from other local researches were used (2, 8). The hourly traffic speed data was not available for all roads, especially for less-travelled roads outside the 5$_{th}$ ring road (Fig. S1). Therefore, the research domain was divided into 27 traffic zones according to intersecting areas of ring roads and administrative districts, and the averaged traffic data of monitoring roads with the same road type and zones were used for non-monitored roads.

The emission factors for various species in this study were from the guide book for on-road emissions published by the Ministry of Ecology and Environment of the People's Republic of China (MEP), most of which were from the bench testing and on-road vehicle emissions measurements in China by Tsinghua University (9). These basic emission factors are significantly different among different vehicle classification. Based on the existing method of MEP, the vehicle can be classified into 8 categories: light duty vehicle (LDV), middle duty vehicle (MDV), heavy duty vehicle (HDV), light duty truck (LDT), middle duty truck (MDT), heavy duty truck (HDT), bus or taxi. According to the fuel type, vehicles are classified into gasoline, diesel, and other (e.g. liquefied natural gas or compressed natural gas). The emission standard is classified as Pre-China I, China I, China II, China III, China IV, or China V. Moreover, the emission factors were corrected by the traffic conditions based on the method listed in the widely used Computer Programme to Calculate Emissions from Road Transport (COPERT) model (10). The $NO_x$ emission was also corrected by the real-world environmental condition collected from the National Climate Data Center (NCDC, ftp://ftp.ncdc.noaa.gov/pub/data/noaa/) integrated surface database, using the method recommended by the Motor Vehicle Emission Simulator (MOVES) model (11).

The vehicle fleet composition data was different for various road types, which was collected by BMCT based on the video data from typical roads in Beijing (Fig. S15). We changed the proportion of HDV to zero during the COVID-19 outbreak and Spring Festival due to two reasons: (1) all inter-provincial passenger transport into or out of Beijing was prohibited to prevent the spread of virus (2) the HDV owned by private companies almost disappeared because of the Spring Festival holiday. The vehicle registration information of each administrative district in Beijing was also used to estimate the proportion of every vehicle category classified by fuel and emission standards.

In the end, the CO, HC, $NO_x$, $PM_{2.5}$, and $PM_{10}$ emissions from vehicle exhausts for each street segments were calculated using the following formula:

$$E_{i,j,p} = L_i \times F_{i,j} \times \sum_t (R_{i,t} \times EF_{t,p})$$

Where $E_{i,j,p}$ was the emission of pollutant $p$ for street link $i$ in $j$ hour, units in g/h; $L_i$ was the length of street link $i$, units in km; $F_{i,j}$ was the total traffic flow for street link $i$ in $j$ hour, units in veh/h; $R_{i,t}$ was the proportion of vehicle category $t$ in total traffic flow for street link $i$; $EF_{t,p}$ was the emission factor of pollutant $p$ for vehicle category $t$, which was influenced by the travel speed and meteorological condition, units in g/km. Also, since vehicular evaporation emissions (VEEs) accounted for about 39% total HC emissions from vehicles, we quantified HC emissions from the running loss, namely the largest contributor among VEEs (12):

$$EV_{i,j} = L_i \times F_{i,j} \times R_{i,gasoline} \times EF_{rl}/V_{i,j}$$

Where $EV_{i,j}$ was the evaporation emissions for street link $i$ in $j$ hour, units in g/h; $L_i$ was the length of street link $i$, units in km; $F_{i,j}$ was the total traffic flow for street link $i$ in $j$ hour, units in veh/h; $R_{i,gasoline}$ was the proportion of gasoline vehicles in total traffic flow for street link $i$; $EF_{rl}$ was the HC emission factor for the running loss process, which was from previous tunnel studies (13-15), units in g/h; $V_{i,j}$ was the average speed of traffic flow for street link $i$ in $j$ hour, units in km/h.

## Section S2 Emissions from other anthropogenic sources

We assembled emissions from multiple recent studies to ensure the accuracy of the emission inventory for the air quality simulation (Table S2). The emissions of all HDT travels which happened in Beijing was calculated using the full-sample enumeration model called TrackATruck, developed in our recent study (16). The HDT positioning data was provided by the governmental HDT monitoring platform. For each travel, TrackATruck mapped the positioning data to multiple trajectories containing information about the operating modes and calculated the emissions along the trajectories. The TrackATruck model can produce extremely high temporal-spatial resolution of day-to-day, hour-to-hour HDT emission maps (0.01° × 0.01°). Based on the HDT emissions in 2018 in our previous study, the year-on-year change in Beijing's road freight volume from 2018 to 2020 was used to estimate the HDT emissions changes in periods that covered the COVID-19 outbreak.

The base emissions from other sources were obtained primarily from the "2+26" Plan, which included urban anthropogenic emissions of 28 cities in the BTH region and its surrounding areas for the year 2018. This emission inventory was developed with the unified method standard and the exhausted source categories at the district or county level. Anthropogenic emissions from other regions in China were collected from the Multi-resolution Emission Inventory for China (MEIC) model developed by Tsinghua University (http://www.meicmodel.org/), with the year of 2017. In addition, emissions from shipping and biomass burning in China were supplemented by Shipping Emission Inventory Model (SEIM) model in our previous work Liu, *et al.* (17) and Cai, *et al.* (18), respectively.

To estimate the emission reduction during the COVID-19 outbreak and Spring Festival period, the activity level for each major sector were investigated respectively, in accordance with the period defined in Fig. 1. Considering the distinctive industrial structure of Beijing compared to the surrounding areas as well as its large proportion of the migrant population, we also estimated the emission reduction separately for Beijing and other regions. It should be noted that, since the emissions of SLOVE model covered only within and surrounding the 6$_{th}$ ring road, the non-HDT vehicle emissions outside this domain were derived from "2+26 Plan" and further applied the reduction rates calculated from the SLOVE model for the Transition and Lockdown periods, respectively.

For emissions from the industry, the reduction rates were estimated based on the power load curve during the Spring Festival (19). The historical year-round power load curve for Beijing in 2018 was collected from National Development and Reform Commission, where a drastic decline appeared during the Spring Festival holiday, beginning from a week before the holiday. According to the monthly electricity consumption data from the National Bureau of Statistics (http://data.stats.gov.cn/), the tertiary industry and residential activities showed constant electricity consumption throughout the winter, while that for the secondary industry showed a significant decrease in February. It could be thus assumed that the drastic decline of the power load curve was mostly related to the reduction of secondary industry activities. Given the statistical data that the electricity consumption from secondary industry accounted for 29% of the whole society (Beijing Statistical Yearbook, 2019), combined with the load curve before and the during the Spring Festival, the reduction rates of electricity consumption from the secondary industry could be estimated – 39% for Transition and 71% for Lockdown period. As the industry of Beijing is dominated by the light industry, we assumed these factors could be used as the proxies for emission reduction.

Residential emissions were mainly composed of cooking smokes from residents and restaurants. Notably, Beijing is a city with a large migrant population, different from the surrounding cities. During the Spring Festival, emissions from residential combustions in Beijing would be reduced as people return to their home towns. Thus, we used the proportion of the migrant population to the local population from the 2019 Beijing Statistical Yearbook (35%) as the reduction rates for residential combustion emissions during the Lockdown period. Their emissions during the Transition period were reduced by half to reflect the gradually decreased population. The cooking emissions from restaurants were assumed as usual until the COVID-19 outbreak. According to the investigation of the China Cuisine Association, 78% of catering enterprises lost more than 100% of their business income during the epidemic period, compared to the last Spring Festival. In this study, we thus assumed their emissions decreased by 70% during the Lockdown period.

Emissions from power industry were assumed constant since there are only a few local power plants in Beijing and a large proportion of its electricity consumption are supported by the surrounding areas. Other sectors were also considered not affected by the COVID-19 and Spring Festival. For other regions outside Beijing, the control factors from Wang et al. (20) were mainly referenced, except the transportation sector, which was obtained from the SLOVE model. Table S1 summarized the emission reduction rates for all anthropogenic sources in Beijing and other regions.

**Section S3 Configurations of meteorological and air quality models**

The models used in this study for basic air quality simulation included the Weather Research and Forecasting Model (WRF, https://www.mmm.ucar.edu/weather-research-and-forecasting-model) with version 3.8.1 and the Community Multiscale Air Quality (CMAQ, https://www.epa.gov/cmaq) model with version 5.2, which were developed by US NCAR (National Center for Atmospheric Research) and US EPA (Environmental Protection Agency), respectively. We applied the WRF-CMAQ modeling system to simulate the air quality in Beijing from January 10th to February 15th in 2020, with 3 days of spin-up time to avoid the influence of the initial condition. As shown in Fig. S16, four nested domains with a horizontal resolution of 36 km×36 km, 12 km×12 km, 4 km×4 km, and 1.33 km×1.33 km, respectively, were used to improve the accuracy of simulated boundary conditions. The first guess field and boundary conditions for WRF were generated from the 6-h NCEP FNL Operational Model Global Tropospheric Analyses dataset. The four-dimensional data assimilation (FDDA) was enabled using the NCEP ADP Global Surface and Upper Air Observational Weather Data (http://rda.ucar.edu). WRF and CMAQ used 34 vertical layers up to 50 hPa. The major physical options in WRF included Morrison 2-Moment microphysics scheme (21), Kain-Fritsch cumulus cloud parameterization (only for domains 01 and 02) (22), the Rapid Radiative Transfer Model (RRTM) longwave and shortwave radiation scheme (23), the Noah land-surface model (24), the TKE scheme Bougeault-Lacarrere (25) and the Mellor–Yamada–Janij´c (MYJ) planet boundary layer scheme (26). The CMAQ model was configured to use the CB05 gas-phase mechanism and the AERO6 aerosol module with aqueous chemistry.

The Single-Layer Urban Canopy Model (SLUCM) coupled with the Noah land surface model was applied in the WRF model to improve the prediction of the meteorological fields. Since SLUCM can represent the urban geometry by assuming infinitely-long street canyons, the thermal and dynamic effects of urban areas are considered, including the shadowing, trapping, and multiple reflections of solar radiation, canopy flows and anthropogenic heat (27-29). Many previous researchers have successfully used it to improve the fine-gridded simulated surface temperature and wind in a big city (30-34). Furthermore, a high-resolution land use data (250m) for 2014 from Finer Resolution Observation and Monitoring-Global Land Cover (FROM-GLC, http://data.ess.tsinghua.edu.cn) was provided for the urban canyon model to define a more realistic underlying surface, especially for urban areas which were reclassified based on the fraction of impervious surface (Fig. S17-a). Considering the heterogeneous spatial distribution of urban morphological characteristics, the gridded Urban Canyon Parameters (UCPs) were also applied in SLUCM, as shown in Fig. S17-b. The basic building data obtained from AMap, including the number of floors and footprint outline for each building, were used to create the UCPs database following the formulations in He*, et al.* (34) and Burian, *et al.* (35).

Considering the non-linear relationship between air pollutant concentrations and emissions, the CMAQ version 5.0.2 with the Integrated Source Apportionment Model (ISAM) was applied to determine the source contribution to the ambient $PM_{2.5}$ and its species concentrations before and during the COVID-19 outbreak (36). The same meteorological field, emissions and configurations were used as described in last paragraphs. In addition, due to limited computational ability, we divided the emissions into five groups to trace them separately in the ISAM model, including emissions from local sources of Beijing (mobiles, industry, domestic and other local sources) and emissions from regional sources outside Beijing. Besides the local mobile source referred to the on-road emissions calculated in this study, other local sectors were classified according to the MEIC emission model. The "other" local sources mainly included emissions from agriculture, off-road traffic and open burning. All chemical components available in the ISAM were tagged, including primary organic matter (POM), element carbon (EC), sulfate, nitrate, ammonium and other nonreactive components. However, due to the limitation of the existing ISAM model, the secondary organic matter (SOM) were not considered in the source apportionment.

Furthermore, the heterogeneous reaction of $SO_2$ was incorporate into CMAQ models with both version 5.2 and 5.0.1 to enhance the sulfate formation, respectively, considering the significant contribution of SIA (secondary inorganic aerosols) formation to fine particles during the heavy $PM_{2.5}$ pollution in northern China, especially for Beijing (37-43). In this study, the chemistry was parameterized using a pseudo-first-order rate constant as following (44, 45):

$$k_i = \frac{\gamma_i v_i A}{4}$$

Where $k_i$ was the heterogeneous rate constant for species $i$ ($s^{-1}$), $A$ was the total aerosol surface area in the Aitken and accumulation mode ($m^2$), $\gamma_i$ was the reactive uptake coefficient, $v_i$ was the thermal velocity (m $s^{-1}$). The uptake coefficients for $SO_2$ heterogeneous reaction, a key parameter to determine the reaction rate, were from Fu, *et al.* (46), which increased rapidly with the growth of ambient relative humidity (RH), especially when RH was higher than 0.5:

$$\gamma_i = \gamma_{dry} \frac{(0.029 + 0.36 RH^{3.7})}{0.29}$$

Where $\gamma_{dry}$ ($6\times10^{-5}$) was the uptake coefficient under the dry condition.

## Section S4 Evaluation of model performances

The WRF model performance was evaluated against ground-level observations in four major meteorological parameters including surface wind speed (WS), surface wind direction (WD), surface temperature (T) and relative humidity (RH). The meteorological observations in Beijing at every 1-h were obtained from the National Climate Data Center (NCDC, ftp://ftp.ncdc.noaa.gov/pub/data/noaa/) integrated surface database. The benchmarks suggested by previous research was used to judge the meteorological performance (the mean biases (MB) ≤ ±0.5 K for T, MB ≤ ±0.5 m/s for WS and MB ≤ ±10° for WD) (47). As shown in Fig. S18 and Table S3, the simulated surface winds are correlated well with observations with MB in the range of related recommendations. High correlation coefficients (R, 0.88, and 0.94) for surface temperature and humidity proved that the model performances were acceptable, although the MB of T was a litter higher than the suggested goal.

We also estimated the model performance of the revised CMAQv5.2 in predicting the $NO_2$, $O_3$, $PM_{2.5}$, and its chemical compositions' concentrations by comparing the modeling results with observations at 34 air quality monitors and three $PM_{2.5}$ composition monitors of Beijing (locations were shown in Fig. S16-b), as described in Table S4. The real-time hourly air quality data (including $NO_2$, $O_3$, and $PM_{2.5}$) based on Thermo Scientific samplers and analyzers, were obtained from the Beijing Municipal Environmental Monitoring Center (BJMEMC, http://zx.bjmemc.com.cn/). In addition, five dominant chemical components of $PM_{2.5}$ including element carbon ($EC$), organic matter ($OM$), sulfate ($SO_4^{2-}$), nitrate ($NO_3^-$) and ammonium ($NH_4^+$) were discussed in detail, and observations were collected from the National Research Program for Key Issues in Air Pollution Control. In general, the model can capture the temporal and spatial variations of the air quality with correlation coefficients (R) higher than 0.5 of all species. The simulated hourly $PM_{2.5}$ was well agreed with observations, with the overall model performance within the performance criteria suggested by Boylan and Russell (48) (Mean fractional bias (MFB) ≤ ±60% and mean fractional error (MFE) ≤ ±75%). The $O_3$ concentrations were slightly underestimated, mainly due to uncertainties in emission inventory and unavoidable deficiencies during meteorological and air quality simulation.

Fig. S19 further investigated the CMAQv5.2 model performances in predicting major $PM_{2.5}$ chemical components, and Table S4 showed their detailed performance statistics. The simulation errors for each species were relatively small, with MBs ranging from 0.3 μg/m3 to 4.3 μg/m3. However, all species were a little overestimated with NMBs (Normalized Mean Bias) ranging from 7.3% to 32.2%, especially for clean days (a day with 24-hour averaged $PM_{2.5}$ concentration lower than 75 μg/m3). Since EC was mainly from the primary emissions, the uncertainties in emissions inventory and meteorological field were probably responsible for this overestimation. During the heavily polluted days, the revised model slightly overestimated the $SO_4^{2-}$ concentration and underestimated the $NO_3^-$ concentration, which can be partly attributed to the selection of uptake coefficients of heterogeneous reactions.

Moreover, we quantified the differences in $PM_{2.5}$ and its species concentration between modeling results from CMAQv5.0.1-ISAM model with the observations (Table S4). In general, the simulation error of $PM_{2.5}$ was also within the recommended criteria (MFB ≤ ±60% and MFE ≤ ±75%), and the performances in predicting EC and SNA (sulfate, nitrate and ammonium) were both similar with those of CMAQv5.2, with a small MBs ranging from 0.3 μg/m3 to 3.6 μg/m3. However, due to the lack of additional SOM formation pathways included in CMAQv5.2 (e.g. aging of S/IVOC and primary organic aerosol), the OM was underpredicted with a NMB of -62.6%, leading to a lower estimated $PM_{2.5}$ concentration (49, 50). As mentioned in section S3, since the SOM formation was not traced in the ISAM model due to the existing limitation, the underestimated SOM would not affect the source apportionment results. In sum, the predicted $PM_{2.5}$ and its chemical components concentrations in both CMAQv5.2 and CMAQv5.0.1-ISAM showed a acceptable agreement with observations, which provided confidence in the source contribution as described in the results section of main text.

**Section S5 Uncertainties**

The uncertainties of this research are mainly from the estimation of emission inventories and the WRF-CMAQ modeling system, as discussed below.

1) Vehicle emission inventories. The on-road emission inventory estimated by the SLOVE model was subject to a few inherent uncertainties and limitations, such as the systematic error of local speed-volume models and the emission factors chosen. In addition, HC emissions from other VEEs processes, including refueling loss, hot soak loss and diurnal loss, were not calculated in this study due to lack of related activity data. However, the traditional uncertainty estimation method (e.g. Monte-Carlo method) was not available for methods based on big data in this study due to large burden on computing time. Therefore, we compared our results with vehicle emissions from other previous studies using a similar bottom-up method (Fig. S6). Our results were within the upper and lower limits of those in literatures, and similar to the emissions for the year of 2018 calculated in "2+26" plan with differences in all pollutants within the range of 8.1%~33.5%, indicating that our vehicle emissions were feasible despite the limitations.
2) Emissions from other sectors. Although the best-available emission inventories of other sectors we obtained were generally developed with the year of 2018 and 2017 (eg. "2+26" Plan, MEIC), there were some changes in emission activities and control measures from the year 2017/2018 to 2020, such as Three-Year Action Plan for Winning the Blue-Sky Defense Battle (51), probably resulting in an overestimation of emissions. However, these uncertainties were relatively minor during the COVID-19 Lockdown period, as most of the industrial activities are largely limited and the emission was reduced to a low level.
3) WRF-CMAQ model. Although we have applied SLUCM in the WRF model and incorporated the heterogeneous reactions of $SO_2$ into the CMAQ model to improve the predictions during haze pollution, there were still inevitable uncertainties due to the chemical processes and physical parameters. In addition, the sensitivity analysis by setting emission reduction scenarios in the WRF-CMAQ model may result in uncertainties of the $O_3$ concentration due to its nonlinear relationship with its gaseous precursors. However, the model performances have been proved to be reliable in Section S4, which was in line with other recent modeling studies in Beijing (52, 53).
4) ISAM model. Due to the limitation of the existing ISAM model, VOC emissions could not be tagged to trace the formation of SOM in the source apportionment. While the SOM was one of the major $PM_{2.5}$ components in winter of Beijing, especially with a rapid growth during heavy pollution days (Fig. 2-A). Therefore, we probably underestimated the contribution of vehicle emissions to $PM_{2.5}$ concentrations since (a) aromatics from gasoline vehicle exhausts is a critical determinant of urban secondary organic aerosol formation (54) and (b) synergetic oxidation of vehicular exhaust leads to efficient formation of ultrafine particles (UFPs) under urban conditions which was also not included in this model (55). Quantification of vehiclar contribution to SOA and UFPs is necessary and suggested for future investigation.

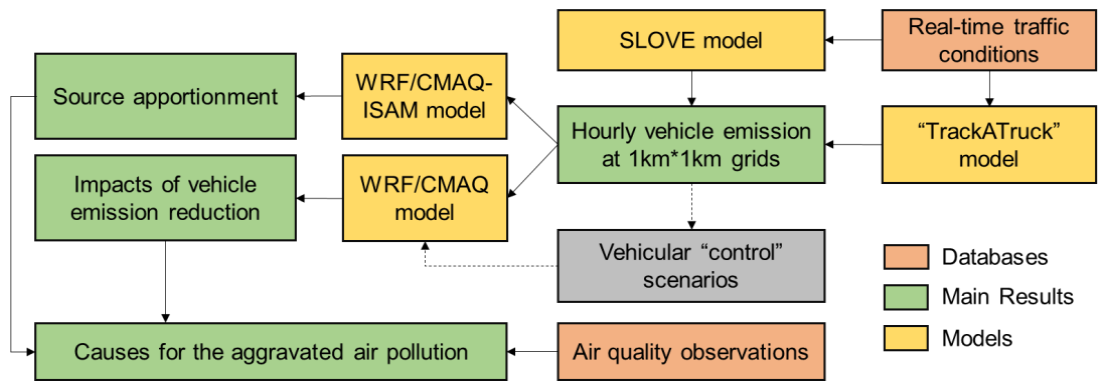

**Fig. S13.** The framework of this research.

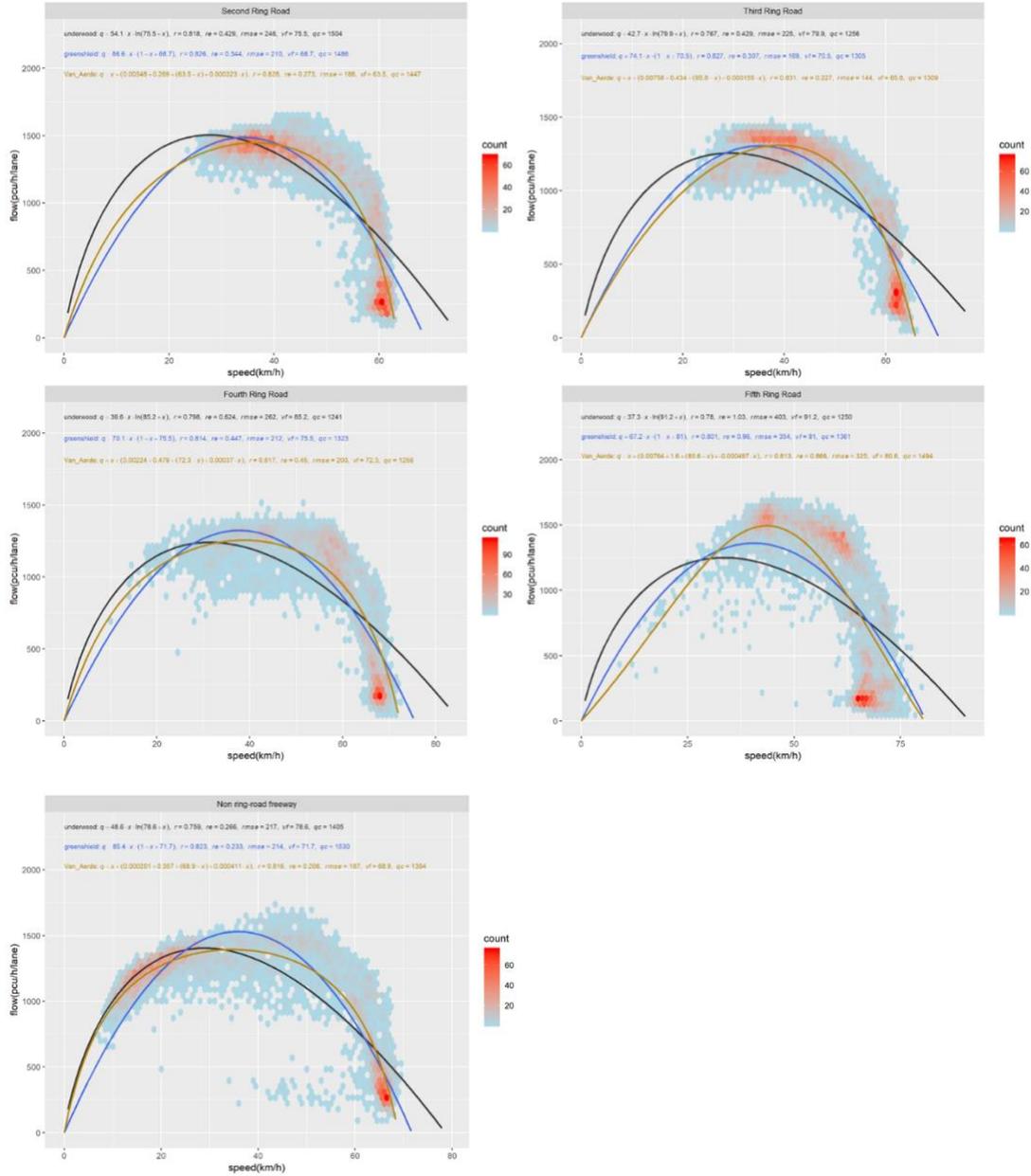

**Fig. S14.** The fitting relationship of traffic speed and volume in different urban freeways of Beijing based on observations. The fitting curves were three typical traffic density models, namely the Underwood model, Greenshield model, and Van Aerde Model. In their formulas, q is the traffic flow, units in pcu·(hour·lane)-1; x is the traffic speed, units in km/h; vf is the free-flow speed, units in km/h; qc is the traffic capacity, units in pcu·(hour·lane). The performance of each model was evaluated using r (correlation coefficient), re (relative error), and RSE (root mean square error).

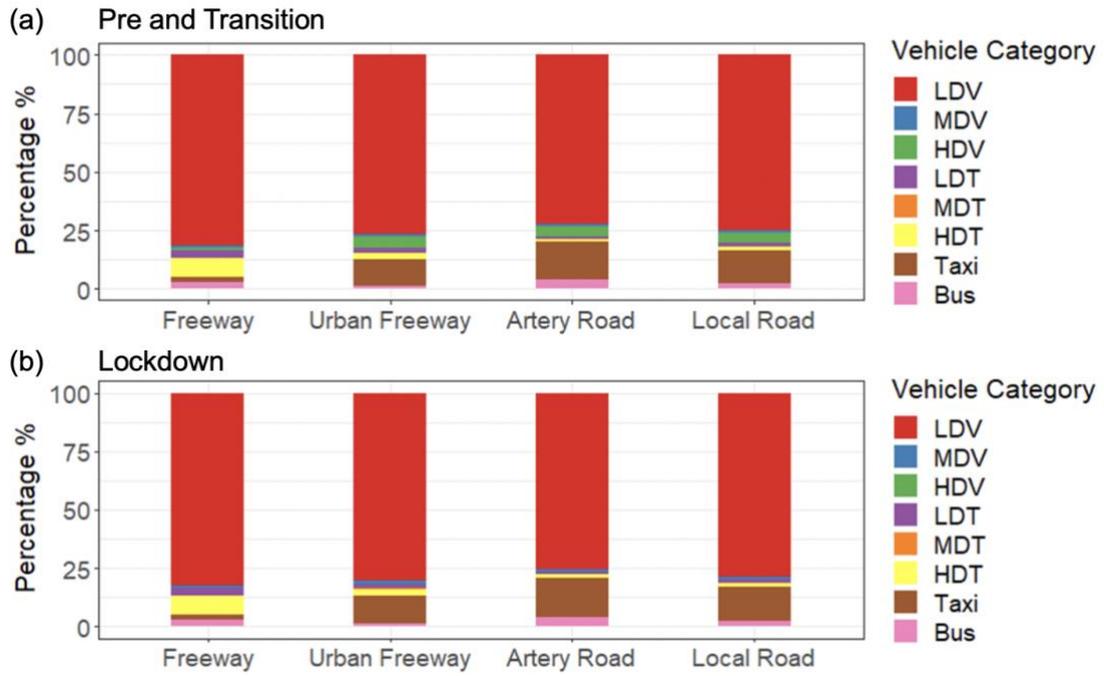

**Fig. S15.** Fleet composition of different road types in Beijing in (a) Pre and Transition period, (b) Lockdown period.

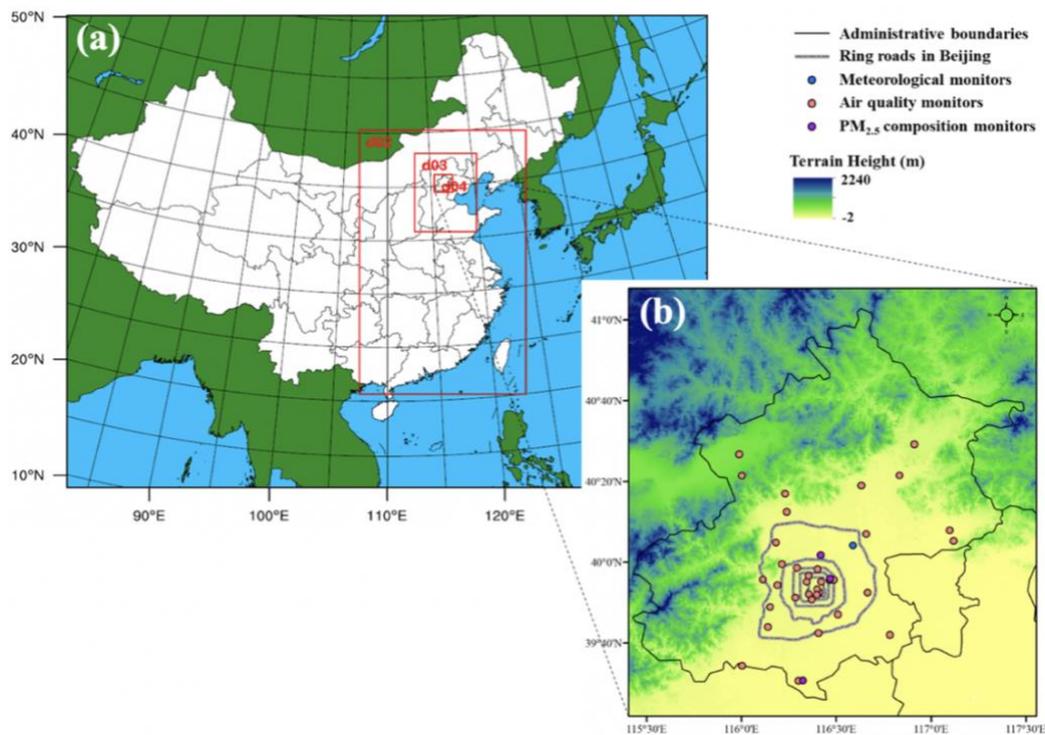

**Fig. S16.** Domains of the WRF-CMAQ modeling system and the location of the monitors. (a) Four nested domains at a horizontal resolution of 36 km×36 km, 12 km×12 km, 4 km×4 km, and 1.33 km×1.33 km, respectively. (b) Meteorological, air quality, and PM$_{2.5}$ composition monitors in Beijing.

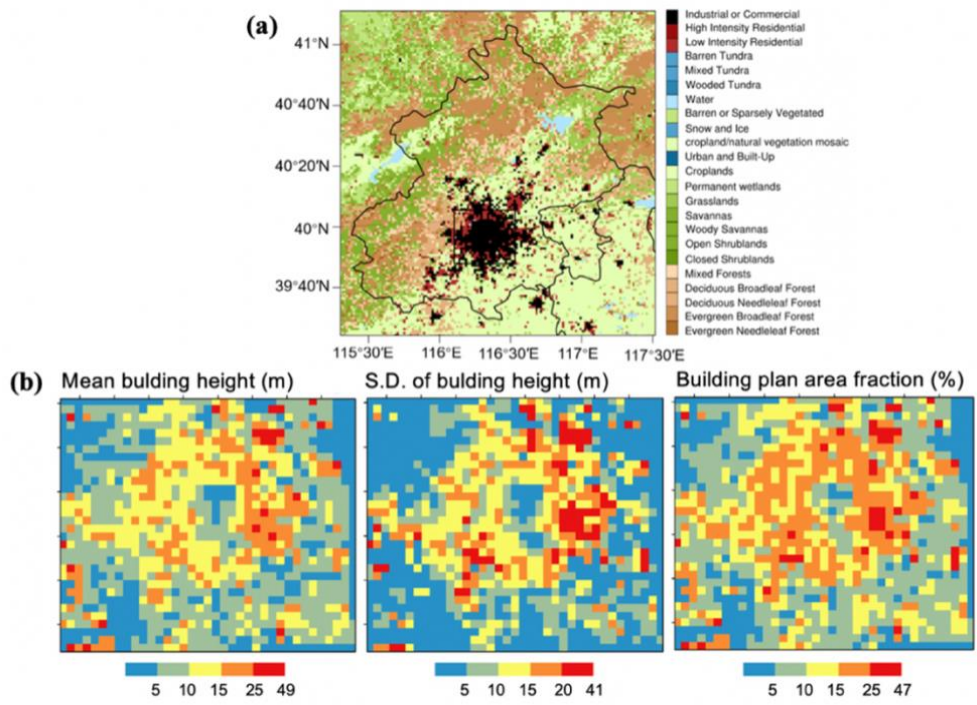

**Fig. S17.** The updated types of land use and UCPs applied in SLUCM coupled with the WRF model. (a) Updated types of land use in Beijing and its surrounding areas from the FROM-GLC database. The black rectangle refers to the domain of gridded UCPs. (b) Three typical UCPs at 1 km×1 km gird in core urban areas of Beijing.

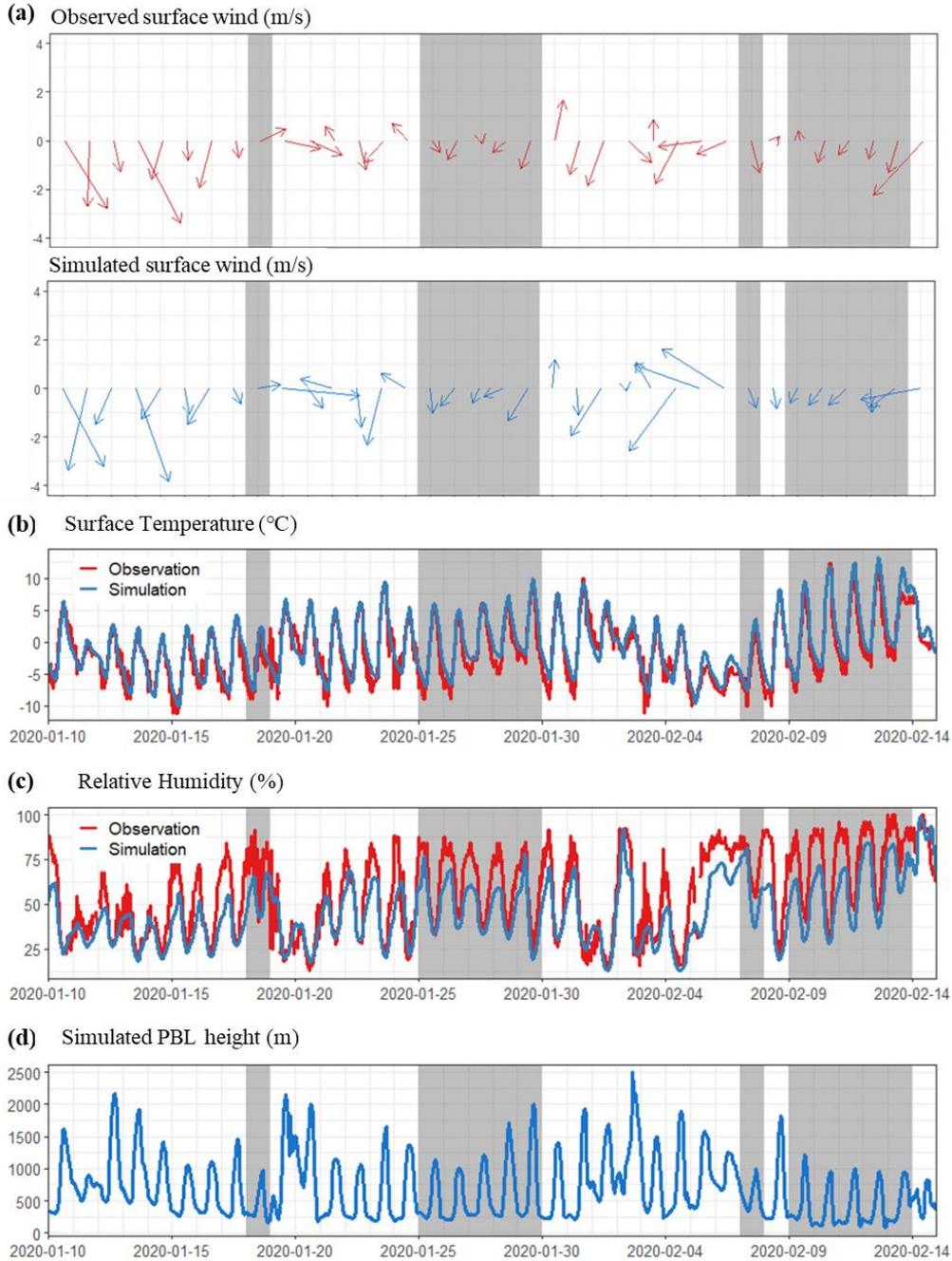

**Fig. S18.** Time series of observed and simulated meteorological parameters including (a) daily surface wind speed and direction calculated from hourly data by vector-averaging, (b) hourly surface temperature, (c) hourly Relative Humidity (RH), (d) hourly simulated Planetary Boundary Layer (PBL) height. The gray background stands for the heavily polluted days when the 24-hour averaged PM$_{2.5}$ concentration exceeds 75 μg/m$_3$.

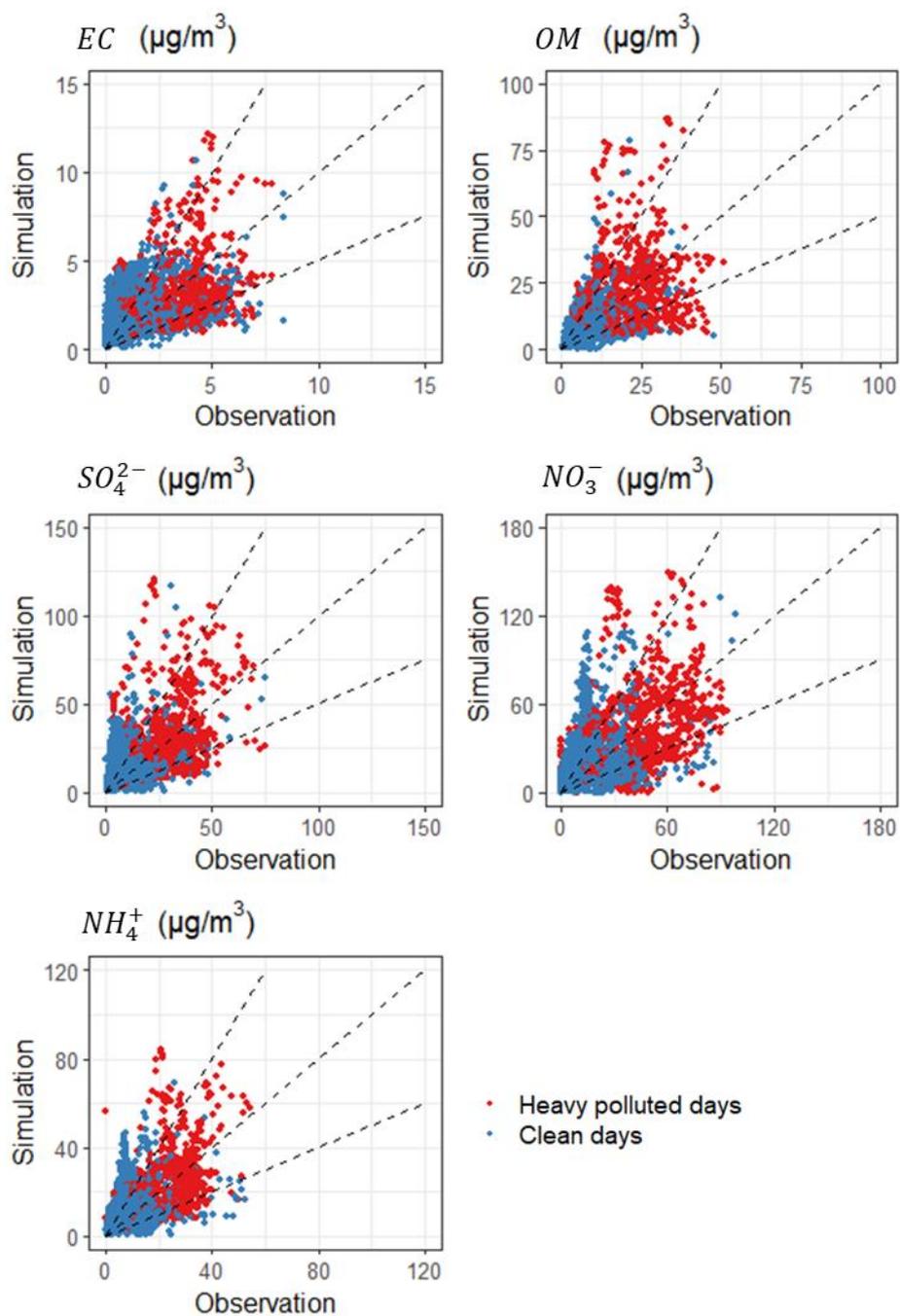

**Fig. S19.** Evaluation of simulated hourly PM$_{2.5}$ chemical composition concentrations against ground-based observations. The dashed lines correspond to the 1:1, 1:2, and 2:1 lines, respectively. Heavily polluted days stand for a day with 24-hour averaged PM$_{2.5}$ concentration higher than 75 μg/m3, and clean days refer to other periods.

**Table S2.** Emissions used in CMAQ model

| Emissions | Base Year | Reference |
|---|---|---|
| On-road emissions of Beijing | Real | This research |
| Urban anthropogenic source emissions of major cities in northern China | 2018 | "2+26" Plan |
| Shipping Emissions | 2013 | SEIM model (17) |
| Biomass burning emissions in China | 2015 | Cai, *et al.* (18) |
| Other anthropogenic source emissions in China | 2017 | MEIC model (http://www.meicmodel.org/) |
| Anthropogenic source emissions outside China | 2010 | MIX emission inventory (56) |
| Emissions from nature sources | Real | Model of Emissions of Gases and Aerosols from Nature (MEGAN) version 3.0 (57) |
| Emissions from windblown dust | Real | In-line calculated in CMAQ (https://www.epa.gov/cmaq) |
| Emissions from sea salt | Real | In-line calculated in CMAQ (https://www.epa.gov/cmaq) |

**Table S3.** Statistics for meteorological predictions.

|  | Sample Number | Mean Observation | Mean Simulation | MB | NMB | RMSE | R |
|---|---|---|---|---|---|---|---|
| **Wind speed (m/s)** | 838 | 2.1 | 2.4 | 0.4 | 17.0 | 1.3 | 0.6 |
| **Wind direction (°)** | 514 | 179.9 | 172.5 | -3.6 | 2.0 | 56.8 | 0.4 |
| **Temperature (°C)** | 857 | -1.3 | -0.5 | 0.7 | -58.6 | 1.8 | 0.9 |
| **Relative humidity (%)** | 856 | 56.7 | 46.0 | -10.7 | 18.8 | 15.5 | 0.9 |

Note: The comparison for wind direction was corrected with the consideration of the periodic nature of wind.

**Table S4.** Model performance statistics for the concentrations of $NO_2$, $O_3$, $PM_{2.5}$, and $PM_{2.5}$ components

| Species (units: μg/m³) | Sample Number | Mean Observation | Mean Simulation | MB | NMB | MFB | MFE | R |
|---|---|---|---|---|---|---|---|---|
| **CMAQv5.2** | | | | | | | | |
| $NO_2$ | 29178 | 34.9 | 36.2 | 1.4 | 2.9 | -2.0 | 53.4 | 0.5 |
| $O_3$ | 29059 | 41.1 | 36.3 | -4.8 | -11.7 | -33.0 | 76.5 | 0.6 |
| $PM_{2.5}$ | 29059 | 79.1 | 72.4 | -6.6 | -7.6 | 2.9 | 55.6 | 0.6 |
| OM | 2550 | 11.2 | 12.1 | 1.0 | 24.5 | 7.3 | 52.1 | 0.6 |
| EC | 2550 | 2.1 | 2.4 | 0.3 | 15.7 | 32.2 | 65.8 | 0.5 |
| $SO_4^-$ | 2535 | 14.3 | 18.7 | 4.2 | 31.5 | 22.5 | 62.8 | 0.5 |
| $NO_3$ | 2535 | 22.3 | 27.0 | 4.3 | 17.3 | 31.7 | 73.8 | 0.6 |
| $NH_4$ | 2519 | 12.3 | 14.6 | 2.0 | 15.6 | 11.7 | 54.0 | 0.6 |
| **CMAQv5.0.1-ISAM** | | | | | | | | |
| $PM_{2.5}$ | 29059 | 79.1 | 61.8 | -17.2 | -21.3 | -8.8 | 58.4 | 0.6 |
| OM | 2550 | 11.2 | 3.4 | -7.7 | -62.6 | -83.5 | 99.9 | 0.3 |
| EC | 2550 | 2.1 | 2.4 | 0.3 | 16.2 | 33.0 | 67.0 | 0.5 |
| $SO_4^-$ | 2535 | 14.3 | 18.1 | 3.6 | 27.1 | 19.6 | 63.1 | 0.5 |
| $NO_3$ | 2535 | 22.3 | 25.1 | 2.5 | 8.9 | 25.3 | 73.5 | 0.5 |
| $NH_4$ | 2519 | 12.3 | 14.0 | 1.3 | 9.7 | 6.7 | 54.8 | 0.6 |

Note: OM, EC, $SO_4^-$, $NO_3$, and $NH_4$ referred to organic matter, element carbon, sulfate, nitrate, and ammonia, respectively. Units of MB (Mean Biases) are μg/m³, and units for NMB (Normalized Mean Bias), MFB (Mean Fractional Bias), MFE (Mean Fractional Error) are %.